\documentclass[12pt]{article}

\usepackage{graphicx}

\usepackage{amsmath}
\usepackage{booktabs}
\usepackage{longtable}

\usepackage{xr}

\usepackage{scicite}
\usepackage{afterpage} 

\usepackage{times}
\usepackage[labelfont=bf,labelsep=period]{caption}

\usepackage[final]{pdfpages}

\usepackage{doi}

\newcounter{firstbib}

\def\sgra{Sgr~A$^{\ast}$}
\def\lsim{\mathrel{\raise.3ex\hbox{$<$\kern-.75em\lower1ex\hbox{$\sim$}}}}
\def\gsim{\mathrel{\raise.3ex\hbox{$>$\kern-.75em\lower1ex\hbox{$\sim$}}}}
\def\gtwid{\mathrel{\raise.3ex\hbox{$>$\kern-.75em\lower1ex\hbox{$\sim$}}}}
\def\proptwid{\mathrel{\raise.3ex\hbox{$\propto$\kern-.75em\lower1ex\hbox{$\sim$}}}}
\DeclareMathOperator{\re}{Re}

\def\aj{Astron. J.} 
\def\araa{Annu.~Rev.~Astron.~Astrophys.}
\def\apj{Astrophys.~J.}
\def\apjl{Astrophys.~J.}
\def\apjs{Astrophys.~J.~Suppl.~Ser.}
\def\aap{Astron.~Astrophys.}
\def\mnras{Mon.~Not.~R.~Astron.~Soc.}
\def\pasj{Publ.~Astron.~Soc.~Jpn.}
\def\nat{Nature}

\newenvironment{sciabstract}{%
\begin{quote} \bf}
{\end{quote}}

\newcounter{lastnote}

\usepackage{authblk}

\title{\bf\LARGE \vspace{-0.5cm} Resolved magnetic-field structure and variability near the event horizon of Sagittarius A*}

\author{\normalsize Michael~D.~Johnson$^{1*}$, Vincent~L.~Fish$^{2}$, Sheperd~S.~Doeleman$^{1,2}$, Daniel~P.~Marrone$^{3}$, Richard~L.~Plambeck$^{4}$, 
John~F.~C.~Wardle$^{5}$, Kazunori~Akiyama$^{6,7}$,  Keiichi~Asada$^{8}$, Christopher~Beaudoin$^{2}$, Lindy~Blackburn$^{1}$, Ray~Blundell$^{1}$, 
Geoffrey~C.~Bower$^{9}$, Christiaan~Brinkerink$^{10}$, Avery~E.~Broderick$^{11,12}$, Roger~Cappallo$^{2}$, Andrew~A.~Chael$^{1}$, Geoffrey~B.~Crew$^{2}$, Jason~Dexter$^{13}$,  
Matt~Dexter$^{4}$, Robert~Freund$^{3}$, Per~Friberg$^{14}$, Roman~Gold$^{15}$, Mark~A.~Gurwell$^{1}$, Paul~T.~P.~Ho$^{8}$, Mareki~Honma$^{6,16}$, Makoto~Inoue$^{8}$, Michael~Kosowsky$^{1,2,5}$,  Thomas~P.~Krichbaum$^{17}$, James~Lamb$^{18}$, Abraham~Loeb$^{1}$, Ru-Sen~Lu$^{2,17}$, David~MacMahon$^{4}$, Jonathan~C.~McKinney$^{15}$, James~M.~Moran$^{1}$, Ramesh~Narayan$^{1}$,  Rurik~A.~Primiani$^{1}$, Dimitrios~Psaltis$^{3}$, Alan~E.~E.~Rogers$^{2}$, Katherine~Rosenfeld$^{1}$, Jason~SooHoo$^{2}$, Remo~P.~J.~Tilanus$^{10,19}$, Michael~Titus$^{2}$,  Laura~Vertatschitsch$^{1}$, Jonathan~Weintroub$^{1}$, Melvyn~Wright$^{4}$, Ken~H.~Young$^{1}$, J.~Anton~Zensus$^{17}$, Lucy~M.~Ziurys$^{3}$\\
\footnotesize{$^{1}$}{Harvard-Smithsonian Center for Astrophysics, 60 Garden Street, Cambridge, MA 02138, USA}\\
\footnotesize{$^{2}$}{Massachusetts Institute of Technology, Haystack Observatory, Route 40, Westford, MA 01886, USA}\\
\footnotesize{$^{3}$}{Steward Observatory, University of Arizona, 933 North Cherry Ave., Tucson, AZ 85721-0065, USA}\\
\footnotesize{$^{4}$}{University of California Berkeley, Dept.\ of Astronomy, Radio Astronomy Laboratory, 501 Campbell, Berkeley, CA 94720-3411, USA}\\
\footnotesize{$^{5}$}{Department of Physics MS-057, Brandeis University, Waltham, MA 02454-0911}\\
\footnotesize{$^{6}$}{National Astronomical Observatory of Japan, Osawa 2-21-1, Mitaka, Tokyo 181-8588, Japan}\\
\footnotesize{$^{7}$}{Department of Astronomy, Graduate School of Science, The University of Tokyo, 7-3-1 Hongo, Bunkyo-ku, Tokyo 113-0033, Japan}\\
\footnotesize{$^{8}$}{Academia Sinica Institute of Astronomy and Astrophysics (ASIAA), P.O.~Box 23-141, Taipei 10617, Taiwan}\\
\footnotesize{$^{9}$}{Academia Sinica Institute for Astronomy and Astrophysics (ASIAA), 645 N.~A`oh\={o}k\={u} Pl., Hilo, HI 96720, USA}\\
\footnotesize{$^{10}$}{Department of Astrophysics/IMAPP, Radboud University Nijmegen, P.O.~Box 9010, 6500 GL Nijmegen, The Netherlands}\\
\footnotesize{$^{11}$}{Perimeter Institute for Theoretical Physics, 31 Caroline Street North, Waterloo, ON N2L 2Y5, Canada}\\
\footnotesize{$^{12}$}{Department of Physics and Astronomy, University of Waterloo, 200 University Avenue West, Waterloo, ON N2L 3G1, Canada}\\
\footnotesize{$^{13}$}{Max Planck Institute for Extraterrestrial Physics, Giessenbachstr. 1, 85748 Garching, Germany}\\
\footnotesize{$^{14}$}{James Clerk Maxwell Telescope, East Asia Observatory, 660 N.~A`oh\={o}k\={u} Pl., University Park, Hilo, HI 96720, USA}\\
\footnotesize{$^{15}$}{Department of Physics, Joint Space-Science Institute, University of Maryland at College Park, Physical Sciences Complex, College Park, MD 20742, USA}\\
\footnotesize{$^{16}$}{Graduate University for Advanced Studies, Mitaka, 2-21-1 Osawa, Mitaka, Tokyo 181-8588}\\
\footnotesize{$^{17}$}{Max-Planck-Institut f\"ur Radioastronomie, Auf dem H\"ugel 69, D-53121 Bonn, Germany}\\
\footnotesize{$^{18}$}{Owens Valley Radio Observatory, California Institute of Technology, 100 Leighton Lane, Big Pine, CA 93513-0968, USA}\\
\footnotesize{$^{19}$}{Leiden Observatory, Leiden University, PO Box 9513, 2300 RA Leiden, The Netherlands}\\
\footnotesize{$^{*}$}{To whom correspondence should be addressed; E-mail:  mjohnson@cfa.harvard.edu.}
}

\renewcommand\Authands{, }

\date{}

\begin{document} 

\baselineskip24pt

\maketitle 


\begin{sciabstract}
Near a black hole, differential rotation of a magnetized accretion disk is thought to produce an instability that amplifies weak magnetic fields, driving accretion and outflow. These magnetic fields would naturally give rise to the observed synchrotron emission in galaxy cores and to the formation of relativistic jets, but no observations to date have been able to resolve the expected horizon-scale magnetic-field structure. We report interferometric observations at 1.3-millimeter wavelength that spatially resolve the linearly polarized emission from the Galactic Center supermassive black hole, Sagittarius A*. We have found evidence for partially ordered fields near the event horizon, on scales of ${\sim}$6 Schwarzschild radii, and we have detected and localized the intra-hour variability associated with these fields.
\end{sciabstract}

Sagittarius A* (\sgra) emits most of its ${\sim}10^{36}$~erg/s luminosity at wavelengths just short of one millimeter, resulting in a distinctive ``submillimeter bump'' in its spectrum \cite{Bower_2015}. A diversity of models attribute this emission to synchrotron radiation from a population of relativistic thermal electrons in the innermost accretion flow \cite{Ozel_2000,Dexter_2010,Yuan_2014}. Such emission is expected to be strongly linearly polarized, ${\sim} 70\%$ in the optically thin limit for a highly ordered magnetic field configuration \cite{Jones_1979}, with its direction tracing the underlying magnetic field. At 1.3-mm wavelength, models of magnetized accretion flows predict linear polarization fractions ${\gsim}30\%$ \cite{Bromley_2001,Yuan_2003,Broderick_Loeb_2006,Shcherbakov_2013}, yet connected-element interferometers measure only a $5{-}10\%$ polarization fraction for \sgra\ \cite{Bower_2005,Marrone_2007}, typical for 
galaxy cores \cite{Jorstad_2005}. 
However, the highest resolutions of these 
instruments, ${\sim}0.1{-}1''$, are insufficient to resolve the millimeter emission region, and linear polarization is not detected from \sgra\ at the longer wavelengths where facility very-long-baseline interferometry (VLBI) instruments offer higher resolution \cite{Bower_1999_2}. Thus, these low polarization fractions could indicate any combination of low intrinsic polarization, depolarization from Faraday rotation or opacity, disordered magnetic fields within the turbulent emitting plasma, or ordered magnetic fields with unresolved structure leading to a low beam-averaged polarization. The higher polarization seen during some near-infrared flares may support the last possibility \cite{Eckart_2006,Zamaninasab_2010}, but the origin and nature of these flares is poorly understood, and they may probe a different emitting electron 
population than is responsible 
for the energetically dominant 
submillimeter emission. 

To definitively study this environment, we are assembling the Event Horizon Telescope (EHT), a global VLBI array operating at 1.3-mm wavelength. Initial studies with the EHT have spatially resolved the ${\sim} 40$~microarcsecond ($\mu$as) emission region of \sgra\ \cite{Doeleman_2008,Fish_2011}, suggesting the potential for polarimetric VLBI with the EHT to resolve its magnetic field structure. For comparison, \sgra\ has a mass of approximately $4.3\times10^6~M_{\odot}$ ($M_{\odot}$, solar mass) and lies at a distance of ${\sim}8~{\rm kpc}$, so its Schwarzschild radius ($R_{\rm Sch} = 2G M/c^2$) is $1.3\times10^{12}$~cm and subtends $10~\mu{\rm as}$  \cite{Ghez_2008,Gillessen_2009}.  
In March 2013, the EHT observed \sgra\ for five nights using sites in California, Arizona, and Hawaii. 
In California, we phased together eight antennas from the Combined Array for Research in Millimeter-wave Astronomy (CARMA) to act as a single dual-polarization station, and we separately recorded dual-polarization data from an additional \mbox{10.4-m} antenna. We also conducted normal observations with CARMA in parallel with the VLBI observations. In Arizona, the 10-m Submillimeter Telescope (SMT) recorded dual polarizations. In Hawaii, seven 6-m dishes of the Submillimeter Array (SMA) were combined into a single-polarization phased array, while the nearby 15-m James Clerk Maxwell Telescope (JCMT) recorded the opposite polarization, forming a single effective dual-polarization station. Each station, except for the CARMA reference antenna, recorded two 512 MHz bands, centered on 229.089 GHz and 229.601 GHz, and circular polarizations. 

A linearly polarized signal manifests itself in the cross-hand correlations between stations, $\left \langle L_1 R_2^\ast \right \rangle$ and $\left \langle R_1 L_2^\ast \right \rangle$, where $L_i$ ($R_i$) denotes left (right) circular polarization at site $i$, and $\ast$ denotes complex conjugation. These correlations are typically much weaker than their parallel-hand counterparts, $\left \langle R_1 R_2^\ast \right \rangle$ and $\left \langle L_1 L_2^\ast \right \rangle$, which measure the total flux. After calibrating for the spurious polarization introduced by instrumental cross-talk \cite{Supplementary_Materials}, quotients of the cross-hand to parallel-hand correlations on each baseline $\textbf{u}$ joining a pair of stations are sensitive to the fractional linear polarization in the visibility domain: $\breve{m}(\textbf{u}) = [\tilde{\mathcal{Q}}(\textbf{u}) + i \tilde{\mathcal{U}}(\textbf{u})]/\tilde{\mathcal{I}}(\textbf{u})$. Here, $\mathcal{I}$, $\mathcal{Q}$, and $\mathcal{U}$ are Stokes 
parameters, and the tilde denotes a 
spatial Fourier transform relating their sky brightness distributions to interferometric visibilities in accordance with the Van~Cittert-Zernike Theorem. 
Importantly, $\breve{m}$ provides robust phase information and is insensitive to station gain fluctuations and to scatter-broadening in the interstellar medium \cite{Johnson_2014,Supplementary_Materials}. 
On baselines that are too short to resolve the source, $\breve{m}$ gives the fractional image-averaged polarization. On longer baselines, $\breve{m}$ mixes information about the spatial distribution of polarization with information about the strength and direction of polarization and must be interpreted with care. For instance, one notable difference from its image-domain analog $m = (\mathcal{Q} + i \mathcal{U})/\mathcal{I}$ is that $|\breve{m}|$ can be arbitrarily large. Nevertheless, $\breve{m}$ readily provides secure inferences about the intrinsic polarization properties of \sgra.

\begin{figure*}[t]
\centering
\includegraphics*[width=1.0\textwidth]{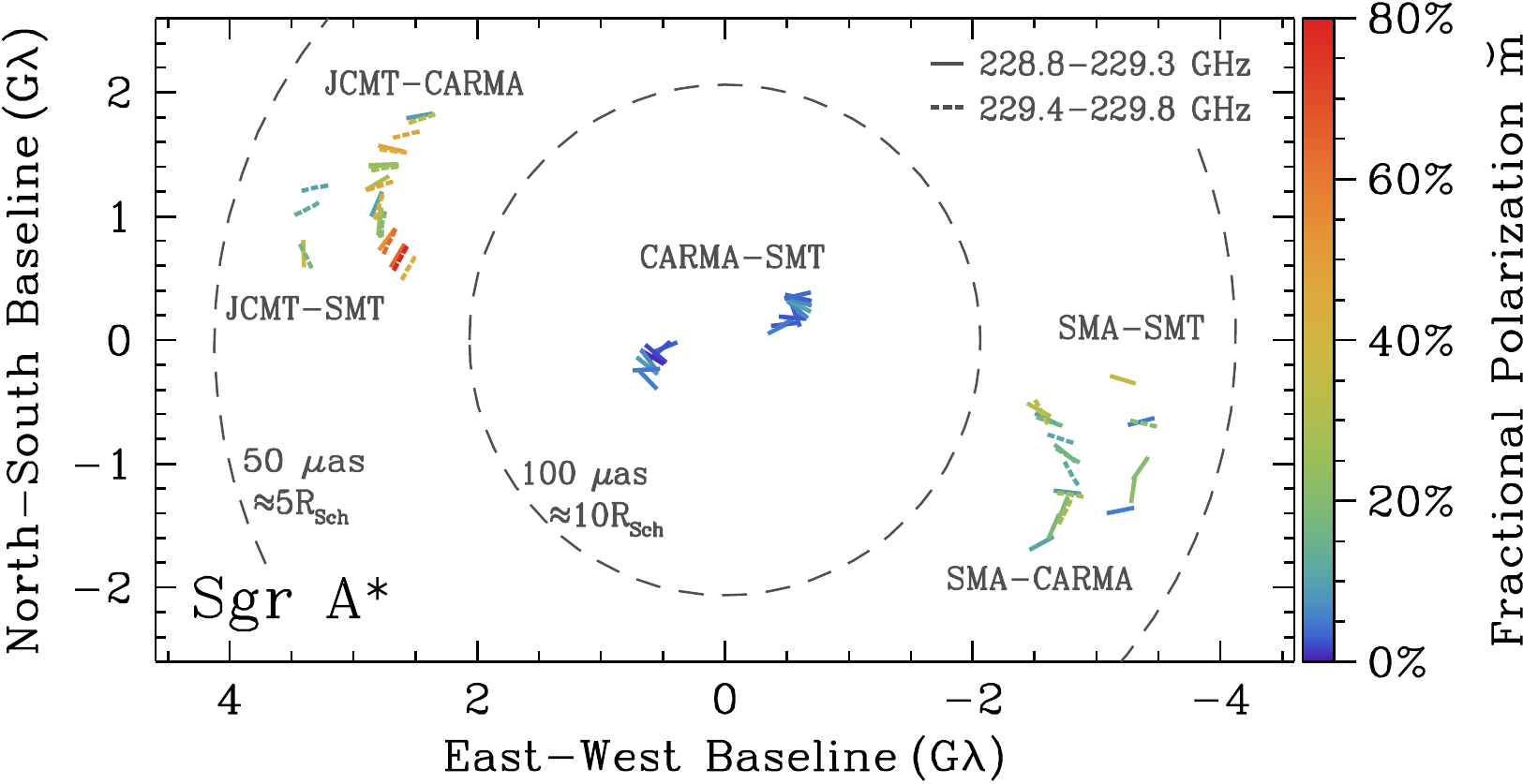}
\caption{ \small
{\bf Interferometric fractional polarization measurements for \sgra.} Interferometric fractional polarization measurements, $\breve{m}(\textbf{u})$, for \sgra\ in our two observing bands during one day of EHT observations in 2013 (day 80). The color and direction of the ticks indicate the (noise-debiased) amplitude and direction of the measured polarization, respectively. For visual clarity, we omit CARMA-only measurements, show only long scans, show only low-band measurements for the SMT-CARMA baseline, and exclude points with a parallel-hand signal-to-noise below 6.5. 
The fractional polarization is expected to change smoothly as the baseline orientation changes with the rotation of the Earth, and the polarization of \sgra\ is also highly variable in time (Fig.~S8). 
The pronounced asymmetry in $\breve{m}(\pm \textbf{u})$ indicates variation in the polarization direction throughout the emission region.
}
\label{fig::SGRA_80_uv}
\end{figure*}

Our measurements on long baselines robustly detect linearly polarized structures in \sgra\ on ${\sim}6 R_{\rm Sch}$ scales (Fig.~1). 
The high (up to ${\sim}70\%$) and smoothly varying polarization fractions are an order of magnitude larger than those seen on shorter baselines  (Fig.~2), suggesting that we are resolving highly polarized structure within the compact emission region. Measurements on shorter baselines show variations that are tightly correlated with those seen in simultaneous CARMA-only measurements (Fig.~S4,S8).  
This agreement demonstrates that there is negligible contribution to either the polarized or total flux on scales exceeding ${\sim}30\, R_{\rm Sch}$, conclusively eliminating dust or other diffuse emission as an important factor \cite{Supplementary_Materials}. Because CARMA does not resolve polarization structure in \sgra, these variations definitively reflect intra-hour intrinsic variability associated with compact structures near the black hole. 

We emphasize that $|\breve{m}(\textbf{u})| {\sim} 70\%$ does not imply correspondingly high image polarizations or that we are measuring polarization near a theoretical maximum. Because polarization can have small-scale structure via changes in its direction, disordered polarization throughout a comparatively smooth total emission region will result in long baselines resolving the total flux more heavily than the polarization \cite{Supplementary_Materials}. As a result, $\left| \breve{m} \right|$ can be arbitrarily large, especially in locations near a visibility ``null'', where $\tilde{\mathcal{I}}$ is close to zero. 
While our highest measured polarization fractions occur where $\tilde{\mathcal{I}}$ falls to only $5-10\%$ of the zero-baseline flux (Fig.~2), the rise is slower than expected for completely unresolved polarization structure, showing that the long baselines are partially resolving coherent polarized structures on the scale of ${\sim}6R_{\rm Sch}$ (Fig.~3). Because interferometric baselines only resolve structure along their direction and our long baselines are predominantly east-west, these conclusions describe the relative coherence of the polarization field in the east-west direction.

\begin{figure}[t]
\centering
\includegraphics*[width=0.5\textwidth]{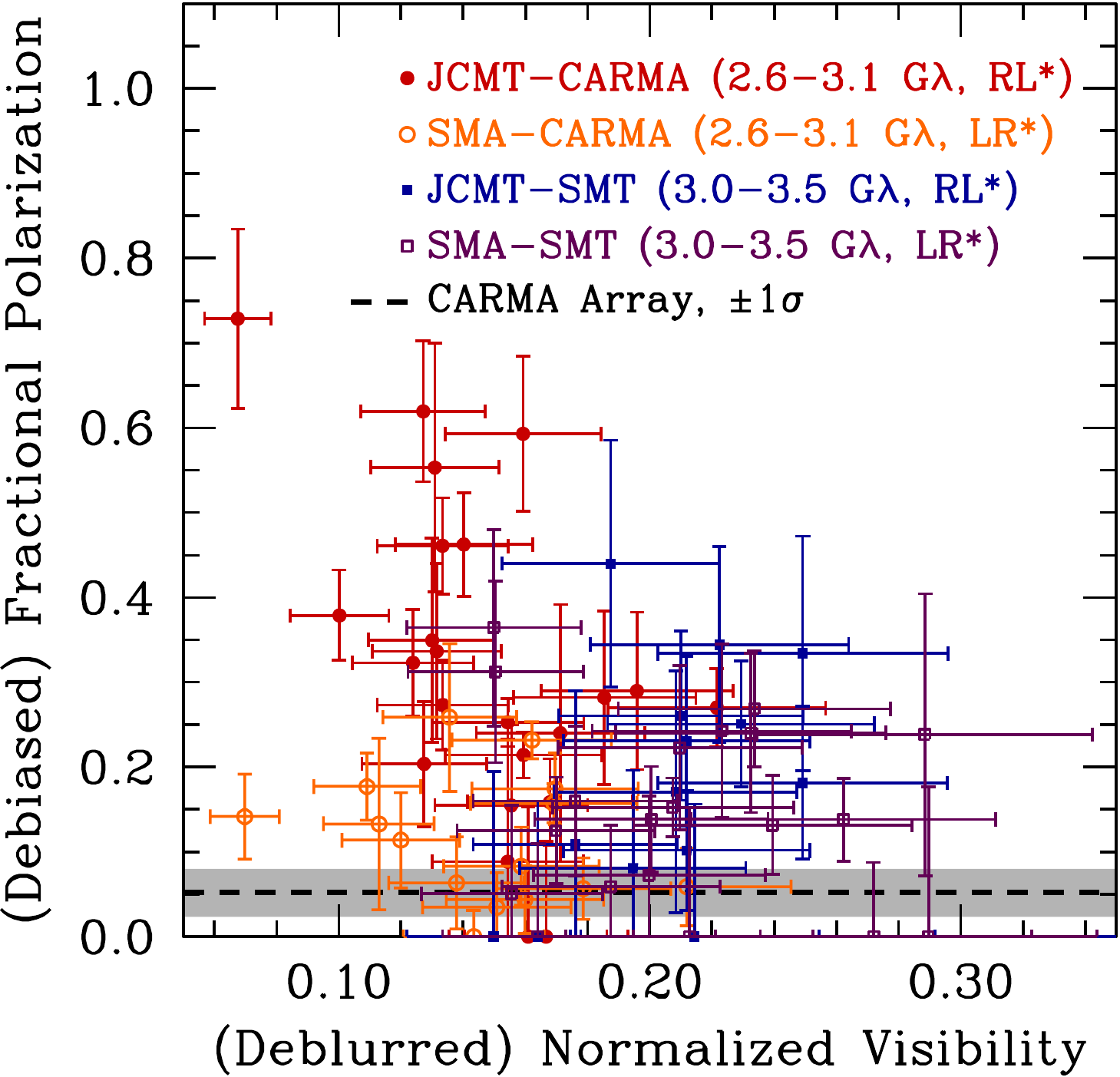}
\caption{\small 
{\bf Signatures of spatially resolved fields from 1.3-mm VLBI.} Long-baseline measurements of interferometric fractional polarization, $\breve{m}$, are plotted against the ``deblurred'' and normalized total-flux visibilities \cite{Fish_2014,Supplementary_Materials}; errors are $\pm 1\sigma$ (refer to Supplementary Materials for details of the error analysis). The black dashed line and gray shaded region show the average and standard deviation of the CARMA measurements of fractional polarization, respectively. The sharp increase in the polarization fraction and variability on the long baselines demonstrates that we are resolving the compact and polarized emission structure on scales of ${\sim}$6-8$R_{\rm Sch}$. 
The marked difference in the two polarization products, $\langle R_1 L_2^\ast \rangle$ and $\langle L_1 R_2^\ast \rangle$, on equal baselines indicates changes in the polarization direction on these scales (cf.~Fig.~1). 
}
\label{fig::CARMA_Long_Compare}
\end{figure}

\begin{figure}[t]
\centering
\includegraphics*[width=0.85\textwidth]{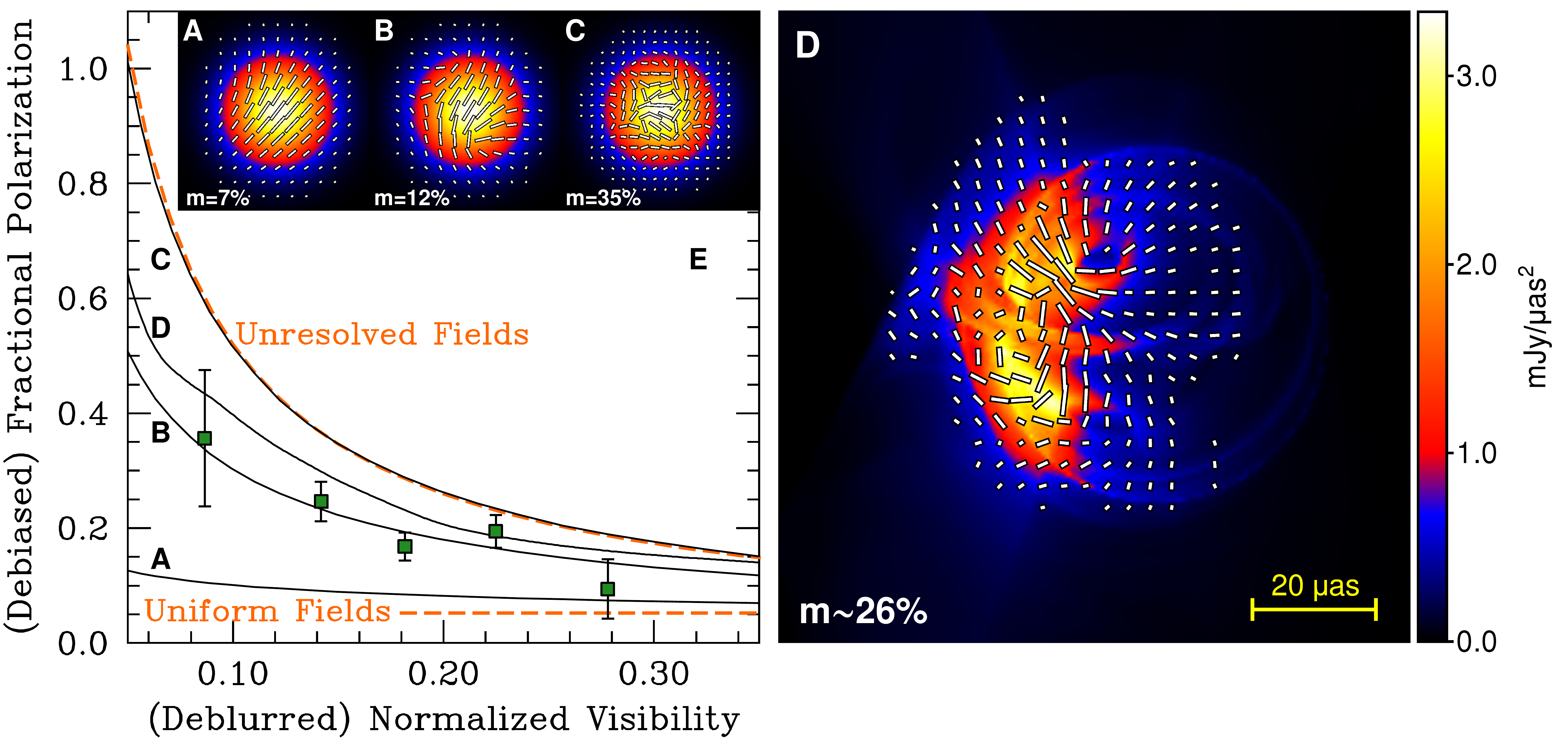}
\caption{\small 
{\bf Strength and order of the polarization field from 1.3-mm VLBI.}
Points with errors ($\pm1\sigma$) in panel \textbf{E} show the 
average of VLBI measurements from Fig.~2 after grouping in bins of 
width 0.05. Dashed orange lines show two limiting cases: a uniform polarization field and a highly disordered (unresolved) polarization field. Each is set equal to 5.2\% when the normalized visibility is unity so that the zero-baseline polarization matches the average of all CARMA-only measurements. Panels \textbf{A}, \textbf{B}, and \textbf{C} show example realizations from three models with Gaussian distributions of intensity. Color shows total flux on a linear scale; ticks show polarization amplitude and direction. Each model has a constant polarization fraction but stochastically varying polarization direction with prescribed coherence lengths (0.64, 0.29, and 0.11 times the Gaussian FWHM). The polarization fractions are determined by matching the ensemble-average zero-baseline polarization to the averaged CARMA measurements. 
Our data differ from model \textbf{A} at a significance exceeding $4\sigma$, differ from model \textbf{C} by $3.4\sigma$, and are compatible with model \textbf{B} \cite{Supplementary_Materials}. 
Panel \textbf{D} shows a sample image from a general relativistic magnetohydrodynamic simulation with polarimetric radiative transfer \cite{Shcherbakov_2013}. The image-averaged polarization fraction, weighted by brightness, is 26\%. This image exhibits a balance between order and variation in the polarization field that is compatible with our observations. 
}
\label{fig::flux_fpol}
\end{figure}

The current data, although too sparse for imaging, provide rich geometrical insights. For instance, if the fractional polarization is constant across the source image, then it will also be constant in the visibility domain. Furthermore, even if the polarization amplitude varies arbitrarily throughout the image, if its direction is constant then the amplitude of the interferometric polarization fraction will be equal for reversed baselines \cite{Supplementary_Materials}. Our measurements (Fig.~1) eliminate both these possibilities and, thus, detect variation in the polarization direction on event-horizon scales.  
These arguments also allow us to assess the spatial extent of the polarized emission because the detected polarization variation cannot arise from a region that is much smaller than the diffraction limit of our interferometer. The polarized emission must therefore span a comparable extent to that of the total flux \cite{Supplementary_Materials}.

The phase of $\breve{m}(\textbf{u})$ likewise constrains the emission morphology. For example, on a short baseline, the leading-order geometrical contribution to interferometric phase comes from the image centroid \cite{Johnson_2014}. 
Thus, just as a decrease in correlated flux with baseline length provides a characteristic angular extent of the emission, a linear change in the phase of $\breve{m}(\textbf{u})$ with baseline length provides a characteristic angular separation of the polarized and total flux. The close agreement in phase between the two measurements on the short SMT-CARMA baseline establishes that the polarized and unpolarized flux are closely aligned, to within ${\sim}10\ \mu{\rm as}$ when the polarization angle of \sgra\ is relatively steady (Fig.~S8). However, 
when variability is dominant, we measure much larger offsets, up to ${\sim}$100~$\mu$as, implicating dynamical activity near the black hole. For comparison, the apparent diameter of the innermost stable circular orbit is $6\sqrt{3/2} R_{\rm Sch} \approx 73\ \mu{\rm as}$, if the black hole of
 \sgra\ is not spinning. The tight spatial association of this linear polarization with the 1.3-mm emission region then cements low-accretion models for \sgra\ \cite{Marrone_2007} which, combined with the measured spectrum in the submillimeter bump and in the near-infrared, imply a magnetic field of tens of Gauss throughout the emitting plasma \cite{Yuan_2003,Dexter_2010,Yuan_2014}.

Even amid magnetically driven instabilities and a turbulent accretion environment, several effects can produce ordered fields near the event horizon. For example, as the orbits of the accreting material around the black hole become circular, magnetic fields will be azimuthally sheared by the differential rotation, resulting in a predominantly toroidal configuration \cite{Hirose_2004}. The high image-averaged polarization associated with the emission region necessitates that such a flow be viewed at high inclination since circular symmetry would cancel the polarization of a disk viewed face-on.  
The striking difference between the stability of compact structures in the total flux \cite{Fish_2011} relative to the rapid changes in the polarized structures on similar scales is then most naturally explained via dynamical magnetic field activity through coupled actions of disk rotation and turbulence driven by the magnetorotational instability (MRI) \cite{Balbus_Hawley_1991}.

Alternatively, accumulation of sufficient magnetic flux near the event horizon may have led to a stable, magnetically dominated inner region, suppressing the disk rotation and the MRI \cite{Narayan_2003,McKinney_2012,Chan_2015}. Emission from a magnetically dominant region provides an attractive explanation for the long-term stability of the circular polarization handedness and the linear polarization direction of \sgra\ \cite{Bower_2002,Munoz_2012}, and it has recently received observational support in describing the cores of active galaxies with prominent jets \cite{Zamaninasab_2014,Mocz_Guo_2014,Marti-Vidal_2015}. However, the close alignment of the polarized and total emission \cite{Supplementary_Materials} severely constrains multi-component emission models for the quiescent flux, such as a bipolar jet \cite{Markoff_2007} or a coupled jet-disk system \cite{Yuan_2002}. If a jet is present, then this constraint suggests substantial differences between the emitting electron populations in 
the jet and the accretion flow to ensure the dominance of a single component at 1.3 mm \cite{Moscibrodzka_2013,Chan_2015}.

With the advent of polarimetric VLBI at 1.3-mm wavelength, we are now resolving the magnetized core of our Galaxy's central engine. 
Our measurements provide direct evidence of ordered magnetic fields in the immediate vicinity of \sgra, firmly grounding decades of theoretical work.  
Despite the extreme compactness of the emission region, we unambiguously localize the linear polarization to the same region and identify spatial variations in the polarization direction. We also detect intra-hour variability and spatially resolve its associated offsets. In the next few years, expansion of the EHT will enable imaging of these magnetic structures and variability studies on the $20$-second gravitational timescale ($G M/c^3$) of \sgra.

\noindent {\bf Acknowledgments\\} EHT research is funded by multiple grants from NSF, by NASA, and by the Gordon and Betty Moore Foundation through a grant to S.D. The Submillimeter Array is a joint project between the Smithsonian Astrophysical Observatory and the Academia Sinica Institute of Astronomy and Astrophysics. The Arizona Radio Observatory is partially supported through the NSF University Radio Observatories program. The James Clerk Maxwell Telescope was operated by the Joint Astronomy Centre on behalf of the Science and Technology Facilities Council of the UK, the Netherlands Organisation for Scientific Research, and the National Research Council of Canada. Funding for CARMA development and operations was supported by NSF and the CARMA partner universities. 
We thank Xilinx for equipment donations. A.B.\ receives financial support from the Perimeter Institute for Theoretical Physics and the Natural Sciences and Engineering Research Council of Canada through a Discovery Grant. Research at Perimeter Institute is supported by the Government of Canada through Industry Canada and by the Province of Ontario through the Ministry of Research and
Innovation. J.D.\ receives support from a Sofja Kovalevskaja award from the Alexander von Humboldt Foundation. M.H.\ was supported by a Japan Society for the Promotion of Science Grant-in-aid. R.T.\ receives support from Netherlands Organisation for Scientific Research. 
Data used in this paper are available in the supplementary materials.

\vspace{1.5ex}

\noindent {\bf Supplementary Materials\\}
www.sciencemag.org\\
Supplementary Text\\
Figs.~S1 to S8\\
Tables S1-S3\\
References (36-61)\\ 
Data S1 and S2

\clearpage

\newcommand{\beginsupplement}{%
        \setcounter{table}{0}
        \renewcommand{\thetable}{S\arabic{table}}%
        \setcounter{figure}{0}
        \renewcommand{\thefigure}{S\arabic{figure}}%
          \setcounter{equation}{0}
        \renewcommand{\theequation}{S\arabic{equation}}
     }

\includepdf[pages=-]{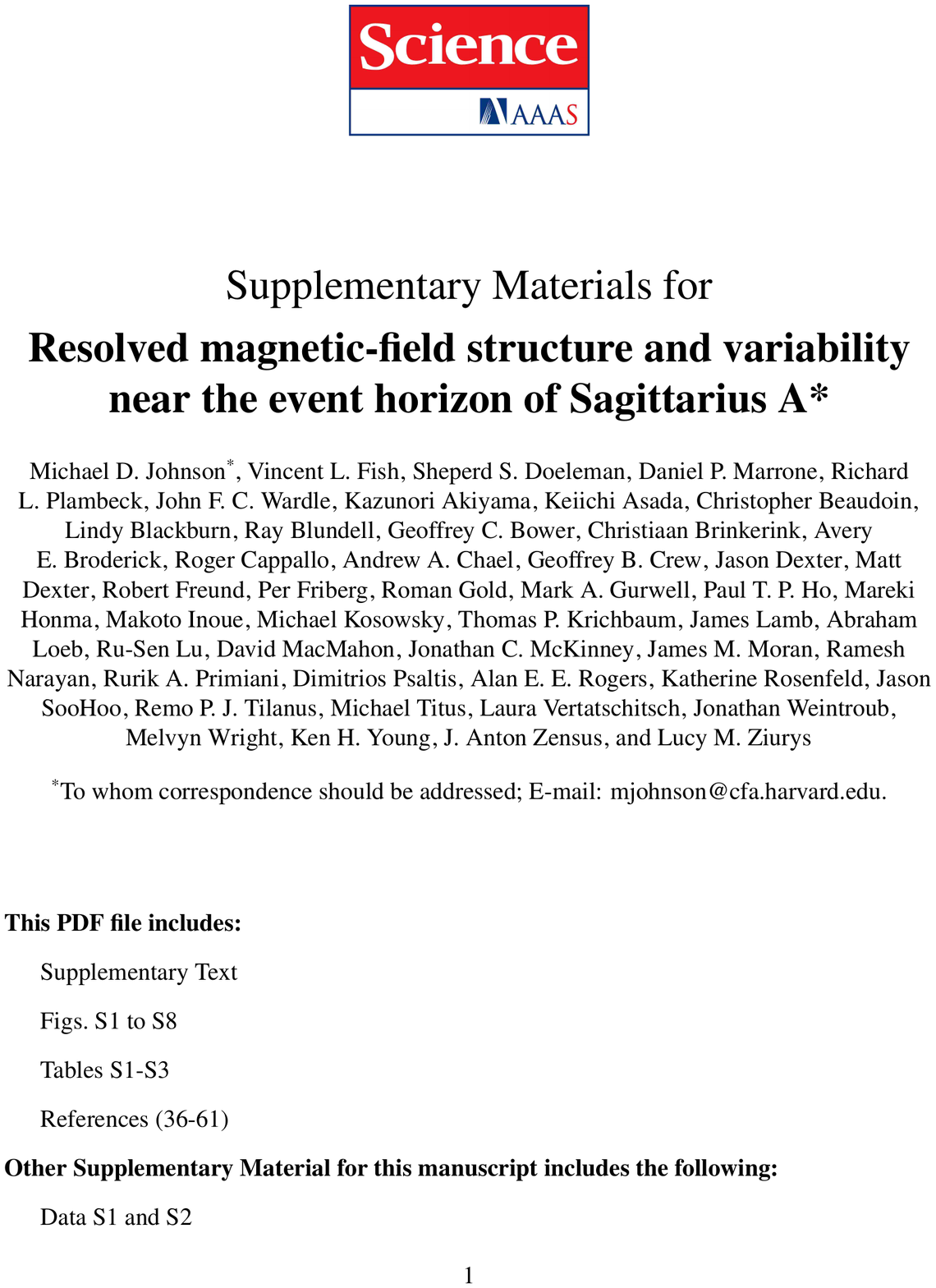}

\beginsupplement

\setcounter{page}{2}

\section*{Supplementary Text}

\section*{Site Details}
\label{sec::Sites}

Polarimetric calibration can be achieved in a general framework, but it depends on specific details about each of the involved stations. For example, we must account for the field rotation angle -- the relative orientation of the feed with respect to a fixed direction on the sky -- which varies with time. The following list provides the specific details needed for each of our sites: 
\begin{itemize}
\item \textbf{CARMA Phased Array:} The CARMA phased array combined antennas C2, C3, C4, C5, C6, C8, C9, and C14. After the second day, the refrigerator dewar in antenna C3 failed; antenna C3 was then replaced by C13 in the phased sum. Sites with an index up to 6 are 10.4-m antennas; the remainder are 6.1-m antennas.
Each telescope recorded both circular polarizations. The phasing efficiency was typically over 90\%, even for weak sources. For each CARMA antenna, the field rotation angle is simply the parallactic angle. As for all sites except the CARMA reference antenna, the CARMA phased array recorded two 512 MHz bands (each with 480 MHz of usable bandwidth). We refer to these as the ``high'' (centered on 229.601 GHz) and the ``low'' (centered on 229.089 GHz) bands. 
\item \textbf{CARMA Reference Station:} CARMA antenna C1 was used separately to record both circular polarizations in the low band only.
\item \textbf{SMT:} The SMT recorded both circular polarizations. The two polarizations have separate Gunn oscillators, resulting in a stochastic phase drift (up to ${\sim}$10$^\circ$/hour) between them. For the SMT, the field rotation angle is the sum of the parallactic and elevation angles.
\item \textbf{Phased SMA:} The phased SMA combined seven of the eight SMA antennas and recorded only one circular polarization at a time. However, because the quarter-wave plates (QWPs) on each antenna can be rapidly rotated, we alternated the recorded polarization: LCP on long scans (4-8 minutes) with interleaved RCP on short scans (30 seconds). 
The estimated phasing efficiency differed between the high and low bands, with medians of 83\% and 65\%, respectively. 
For the SMA, the field rotation angle is $45^\circ$ minus the elevation angle plus the parallactic angle.
\item \textbf{JCMT:} The JCMT recorded a single polarization (RCP), and its field rotation angle is the parallactic angle.
\end{itemize}

\section*{Correlation Procedure}
\label{sec::Correlation}

We correlated our data using the Haystack Mark4 correlator \cite{Whitney_2004}. For subsequent fringe-fitting, we used the Haystack Observatory Post-processing System. Because a typical atmospheric coherence time (${\sim}5$ seconds) is much shorter than our scan lengths (${\sim}5$ minutes), we performed fringe-fitting in three steps \cite{RWB94,Wardle97}:
\begin{enumerate}
\item First, a blind fringe search determined the fringe-rate and the fringe-delay, as well as the RCP-LCP delay differences for dual-polarization stations, which were nearly constant each day. Concurrently, this pass estimated the time-dependent phase, the ``ad hoc'' phase, on parallel-hand baselines to the CARMA phased array for one chosen polarization. This phase included contributions from the source, the atmosphere, the instrument, and thermal noise; the most rapid changes were from the atmosphere and noise. To preserve the polarimetric identities (described below), there was necessarily a single ad hoc phase for both polarizations. Hence, for the ad hoc phases, the JCMT and the SMA were considered to be a single station.  
\item A second pass removed the ad hoc phases calculated in the first pass before fringe fitting. When possible, the removed ad hoc phases were determined from the opposite band (low or high) to avoid biasing weak parallel-hand detections by removing the phases associated with their own thermal noise.  
\item A third and final pass used the delay and rate solutions from the second pass together with the RCP-LCP delay offsets and ad hoc phases from the first pass to coherently average parallel- and cross-hand visibilities over scans.
\end{enumerate}

\section*{Polarimetric Calibration Procedure}
\label{sec::PolarimetricCalibration}

\paragraph*{Definitions and Conventions.} 
\label{sec::Conventions}
Polarimetric VLBI enables studies of the images of the Stokes parameters $\mathcal{Q}$, $\mathcal{U}$, and $\mathcal{V}$ in addition to that of total intensity, $\mathcal{I}$. Polarimetric interferometry on a baseline $\textbf{u}$ (length in wavelengths) is sensitive to the respective Fourier components of these images; e.g., $\tilde{\mathcal{Q}}(\textbf{u})$ \cite{RWB94,TMS}. Moreover, as we have mentioned in the main text and discuss in more depth subsequently, fractional polarization in the visibility domain provides an especially robust observable: $\breve{m}(\textbf{u}) \equiv \left[\tilde{\mathcal{Q}}(\textbf{u}) + i \tilde{\mathcal{U}}(\textbf{u})\right]/\tilde{\mathcal{I}}(\textbf{u})$. Note that we use a breve rather than a tilde to emphasize that this interferometric fractional polarization 
holds no Fourier relationship with its image-domain analog, $m(\textbf{x}) \equiv \left[{\mathcal{Q}}(\textbf{x}) + i {\mathcal{U}}(\textbf{x})\right]/\mathcal{I}(\textbf{x})$. Also, fractional polarization in the Fourier domain is not constrained to have amplitude less than unity, and $\breve{m}(\textbf{u})$ includes contributions to phase from the geometrical distribution of polarized and total flux (e.g., in the phase of $\tilde{\mathcal{I}}(\textbf{u})$). 
The linear polarization, $\mathcal{Q} + i\mathcal{U}$, is commonly written in the form $\mathcal{I} |m| e^{2i \chi}$, where $0 \leq |m| \leq 1$ is the  polarization fraction and $\chi$ is the polarization direction (or angle). We use the same terminology to refer to the analogous quantities for $\breve{m}$, although (as we have noted) these must be interpreted with care.

Imperfections in the instrumental response distort the relationship between the measured polarization and the source polarization. These imperfections can be conveniently described by a Jones matrix, $\textbf{J}$; estimates of the Jones matrix can then be used to correct the distortion. The Jones matrix is commonly decomposed into ``gain'' and ``leakage'' terms, $G_{\rm R,L}$ and $D_{\rm R,L}$, respectively. Specifically, after accounting for field rotation by an angle $\phi$ relative to a fixed (celestial) basis, the measured fields $E_{\rm R,L}'$ can be written
\begin{align}
 \begin{pmatrix}
  E_{\rm R}' \\
  E_{\rm L}'
 \end{pmatrix}
 &\approx
 \begin{pmatrix}
  G_{\rm R} & 0 \\
  0 & G_{\rm L}
 \end{pmatrix}
 \begin{pmatrix}
  1 & D_{\rm R} \\
  D_{\rm L} & 1
 \end{pmatrix}
  \begin{pmatrix}
  e^{-i\phi} & 0 \\
  0 & e^{i \phi}
 \end{pmatrix}
 \begin{pmatrix}
  E_{\rm R} \\
  E_{\rm L}
 \end{pmatrix}
 \equiv 
 \textbf{J}
  \begin{pmatrix}
  E_{\rm R} e^{-i \phi} \\
  E_{\rm L} e^{i \phi}
 \end{pmatrix}
 .
\end{align}
When transformed from a circular-polarization basis to a linear-polarization basis, $\phi$ is simply the angle in a $2\times2$ rotation matrix.

Linearizing in the source polarization and leakage terms and neglecting circular polarization yields the following approximations \cite{RWB94}: 
\begin{align}
\label{eq::PolarimetricEquations}
\frac{\left \langle L_1 R_2^\ast \right \rangle}{\left \langle L_1 L_2^\ast \right \rangle} &\approx \left(\frac{G_{2,\rm R}}{G_{2,\rm L}} \right)^\ast \left[ \breve{m}^\ast(-\textbf{u}_{12}) e^{2i \phi_2} + D_{1,\rm L} e^{-2i (\phi_1-\phi_2)} + D_{2,\rm R}^\ast \right]\\
\nonumber \frac{\left \langle L_1 R_2^\ast\right \rangle}{\left \langle R_1 R_2^\ast\right \rangle} &\approx \left(\frac{G_{1,\rm L}}{G_{1,\rm R}} \right) \left[ \breve{m}^\ast(-\textbf{u}_{12}) e^{2i \phi_1} + D_{1,\rm L} + D_{2,\rm R}^\ast e^{2i (\phi_1-\phi_2)} \right] \\
\nonumber \frac{\left \langle R_1 L_2^\ast\right \rangle}{\left \langle L_1 L_2^\ast\right \rangle} &\approx \left(\frac{G_{1,\rm R}}{G_{1,\rm L}} \right) \left[ \breve{m}(\textbf{u}_{12}) e^{-2i \phi_1} + D_{1,\rm R}  + D_{2,\rm L}^\ast e^{-2i (\phi_1-\phi_2)} \right] \\
\nonumber \frac{\left \langle R_1 L_2^\ast\right \rangle}{\left \langle R_1 R_2^\ast\right \rangle} &\approx \left(\frac{G_{2,\rm L}}{G_{2,\rm R}} \right)^\ast \left[ \breve{m}(\textbf{u}_{12}) e^{-2i \phi_2} + D_{1,\rm R} e^{2i (\phi_1-\phi_2)} + D_{2,\rm L}^\ast \right] \\
\nonumber \frac{\left \langle R_1 R_2^\ast\right \rangle}{\left \langle L_1 L_2^\ast\right \rangle} &\approx \left(\frac{G_{1,\rm R}}{G_{1,\rm L}} \right)\left(\frac{G_{2,\rm R}}{G_{2,\rm L}} \right)^\ast e^{-2i (\phi_1 - \phi_2)}.
\end{align}
Here, $\breve{m}(\textbf{u}_{12})$ represents the fractional polarization in the visibility domain on a baseline joining sites 1 and 2, $\phi_{i}$ is the field rotation angle at site $i$, and we have introduced the simplified notation $R_i \equiv E_{\rm R,i}'$. Note that to recover the full information $\breve{m}(\pm \textbf{u})$ on a given baseline, both polarizations must be recorded at each site.  

\begin{figure}[t]
\centering
\includegraphics*[width=1.0\textwidth]{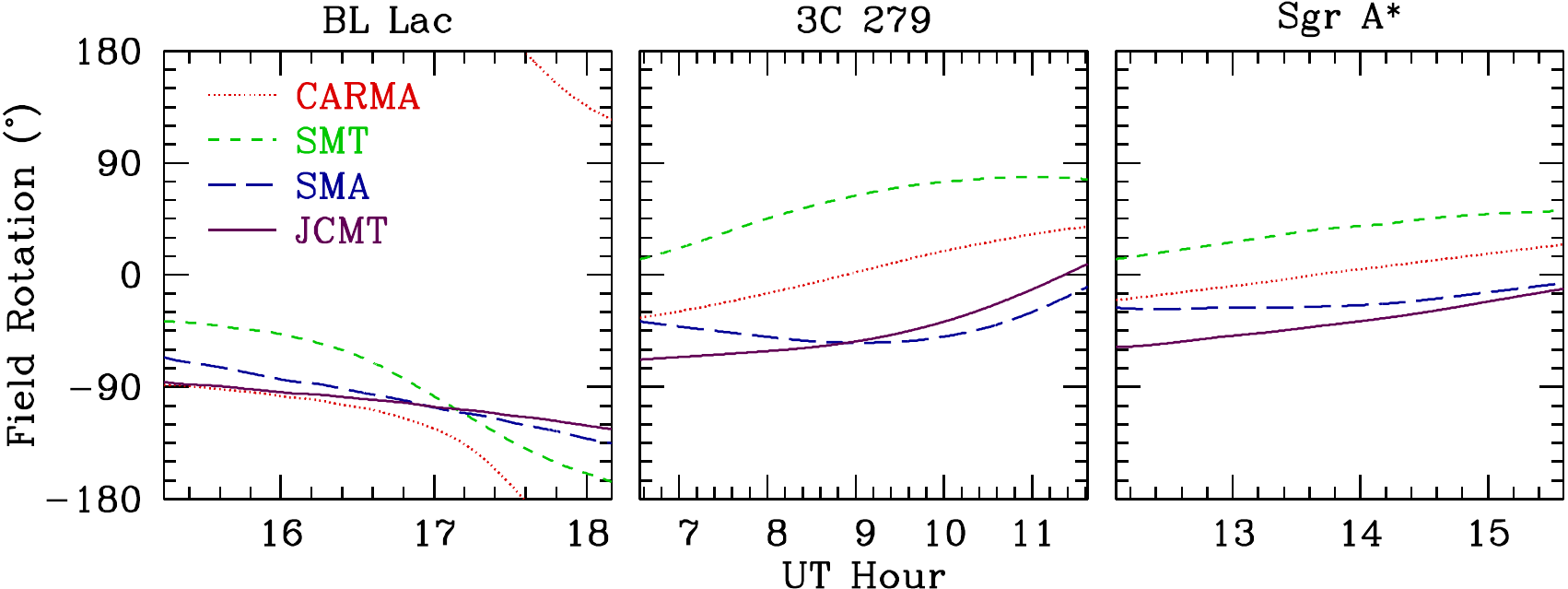}
\caption{ \small 
{\bf Field rotation angles at each EHT site.} Field rotation angles $\phi_i$ at each EHT site as a function of time for our two primary calibrators, BL Lac and 3C~279, and for \sgra. The spurious linear polarization from leakage rotates by {twice} the field rotation angle relative to the true source polarization (Eq.~\ref{eq::PolarimetricEquations}).
}
\label{fig::FieldRotation}
\end{figure}

These equations demonstrate the utility of fractional polarization in VLBI. Namely, variations in the complex gains only affect these quotients through their changes relative to the gain of the opposite circular polarization at the same site. Also, the leakage terms introduce spurious linear polarizations, which differ from the source polarization terms in their dependence on field rotation. Hence, polarimetric calibration relies heavily upon long calibration tracks with wide sampling of the field rotation. Fig.~\ref{fig::FieldRotation} shows the field rotation angles for our two primary calibration targets and for \sgra.

It is straightforward to estimate the gain terms. For example, in the linear approximation of Eq.~\ref{eq::PolarimetricEquations}, parallel-hand ratios on a triangle of stations determine the amplitude of the gain ratios directly: 
\begin{align}
\label{eq::GainRatio}
\left| \frac{G_{1,\rm R}}{G_{1,\rm L}} \right| = 
\sqrt{
\frac{\left \langle R_1 R_2^\ast\right \rangle}{\left \langle L_1 L_2^\ast\right \rangle} \times
\frac{\left \langle R_3 R_1^\ast\right \rangle}{\left \langle L_3 L_1^\ast\right \rangle} \times 
\frac{\left \langle L_3 L_2^\ast\right \rangle}{\left \langle R_3 R_2^\ast\right \rangle}.
}
\end{align}
Because gain terms are only relevant for sites that record both polarizations simultaneously, the set of parallel-hand quotients determines all of the complex gain ratios, up to a {single} unknown phase. This final phase is provided by the absolute electric vector position angle (EVPA) calibration of the array -- in our case, by the complementary EVPA measurements by CARMA (discussed in detail below). Hence, the major challenge in polarimetric calibration is to determine the leakage terms.

\paragraph*{Estimating the Bias and Uncertainty in Interferometric Fractional Polarization.} 
We now describe the error analysis for our estimates of $\breve{m}$. 
To excellent approximation, the cross- and parallel-hand visibilities are drawn from complex Gaussian distributions with known standard deviations. A measured fractional polarization, denoted with a prime ($'$), can then be related to the true fractional polarization as 
\begin{align}
\breve{m}' = \breve{m} \left( \frac{1 + \epsilon_{\rm P}/S_{\rm P}}{1 + \epsilon_{\rm I}/S_{\rm I}} \right),
\end{align}
where $S_{\rm P}$ is the signal-to-noise ratio of the measured $\tilde{\mathcal{P}}$, $S_{\rm I}$ is the signal-to-noise ratio of the measured $\tilde{\mathcal{I}}$, and the $\epsilon$ are complex Gaussian random variables with a standard deviation of unity in their real and imaginary parts. 

Because we rely on strong parallel-hand detections for fringes, $S_{\rm I} \gg 1$ for all of our measurements. Thus,
\begin{align}
\breve{m}' \approx \breve{m} \left( 1 + \epsilon_{\rm P}/S_{\rm P} \right) \left( 1 - \epsilon_{\rm I}/S_{\rm I}\right).
\end{align}
In addition, nearly all of our measurements have $S_{\rm P} \gsim 1$, so $\breve{m}'$ is approximately a complex Gaussian random variable centered on $\breve{m}$ with a signal-to-noise of $S_{\rm m} \approx 1/\sqrt{1/S_{\rm I}^2+1/S_{\rm P}^2}$. [This approximation is best for our measurements with the highest fractional polarizations because they have $S_{\rm P} \gg 1$.] A gain correction will scale $\breve{m}'$, keeping the signal-to-noise ratio $S_{\rm m}$ constant, while a leakage correction will shift $\breve{m}'$, keeping the root-mean-square noise $|\breve{m}'|/S_{\rm m}$ constant but changing the signal-to-noise ratio (see Eq.~\ref{eq::PolarimetricEquations}).

Because of thermal noise, estimates $A'=|\breve{m}'|$ of the fractional polarization amplitude $|\breve{m}|$ will be biased upward. To derive an unbiased estimator $A$ of $|\breve{m}|$, note that $A$ will have a standard deviation of $\sigma_{\rm A} \approx A/S_{\rm m}$ \cite{TMS}. As a result, $\left \langle A^2 \right \rangle = |\breve{m}|^2 + \sigma_{\rm A}^2$. Since $\left \langle A'^2 \right \rangle = |\breve{m}|^2 + 2\sigma_{\rm A}^2$, an appropriate unbiased estimator of $|\breve{m}|$ is $A = \sqrt{ A'^2 - \sigma_{\rm A}^2 }$ (imaginary values are set equal to zero). This standard result for polarimetry was rigorously derived in \cite{Wardle_Kronberg_1974} via a maximum-likelihood approach.  

Interestingly, this estimator differs from the estimator for incoherent averaging of $M$ visibilities: $A_{\rm inc} = \sqrt{\langle A_i'^2 \rangle_M - 2\sigma_{\rm A}^2}$ \cite{RDM95}, where $\langle \dots \rangle_M$ denotes an average of $M$ samples $\{ A_i' \}$. The reason for the difference is that the derivation of $A_{\rm inc}$ implicitly assumes $M \rightarrow \infty$ so that $A_{\rm inc}$ has no associated noise. For a finite number of samples, each having noise $\sigma_{\rm A} \lsim A$, the amplitude estimator will instead have a standard deviation of $\sigma_{\rm A}/\sqrt{M}$. In this case, an unbiased estimator that is appropriate for an arbitrary number of samples is $A = \sqrt{\langle A_i'^2 \rangle_M - \sigma_{\rm A}^2 (2 - 1/M )}$. Taking $M=1$ and $M\rightarrow \infty$ then recovers the standard results for polarization and incoherent averaging, respectively. 

Although these results assume high signal-to-noise for both $\tilde{\mathcal{I}}$ and $\tilde{\mathcal{P}}$, they are excellent for our data. For instance, with $S_{\rm I} = 8$ and $S_{\rm P}=3$, the above prescription to estimate $|\breve{m}|$ has a fractional bias less than $1\%$ and correctly estimates the signal-to-noise ratio to within 10\%. Even when $S_{\rm I} = 6.5$ (the cutoff we have used) and $S_{\rm P}=2$, the fractional bias is less than 2\% and the noise is estimated correctly to within $12\%$. Furthermore, while the quotient of Gaussian random variables is not Gaussian and results in a high tail at large fractional polarizations, the consistency in our results between different frequency bands demonstrates that our measured high polarization fractions are not statistical outliers.

\paragraph*{Data Subset and Preprocessing for Polarimetric Calibration.} 
\label{sec::Preprocessing} 
We performed minimal processing on the visibilities before deriving our calibration solution. Specifically, we only included fractional polarizations for which the parallel-hand visibility had a signal-to-noise ratio (SNR) exceeding 8. We then used the parallel-hand ratios to estimate the RCP-LCP phase drift at the SMT by assuming that the RCP-LCP phase at CARMA was constant on each day. We performed a weighted (by SNR) average of the five nearest phase estimates for each time to invert the phase drift. For calibration, we did not include short (30 second) scans on any baseline except those to the SMA (RCP). We derived our calibration solutions using only 3C~273, 3C~279, and BL Lac. These sources were all bright and exceptionally compact. 

For the purposes of calibration, we also included data on the ${\sim}$200-m JCMT-SMA baseline. Although information on this baseline was redundant with that of CARMA, polarization measurements on the JCMT-SMA baseline provided excellent information about the leakages at these two stations, especially because their field rotation angles differed by the source elevation angle minus $45^\circ$. However, because the JCMT and the SMA each recorded a single polarization, we could not compute the instantaneous fractional polarization on this short baseline.  Nevertheless, we estimated fractional polarization using non-simultaneous estimates of the cross- and parallel-hand visibilities. The parallel-hand products were estimated using the short (30 second) scans when the SMA recorded RCP, and the cross-hand products were estimated using the immediately following long (${\sim}5$ minute) scans. 
Although the phase accuracy of these fractional polarizations is unreliable, the amplitude is robust and varies as the field rotation angles changes at each site because of the changing contribution of the leakage terms.

\paragraph*{Complementary Measurements with CARMA.} 
\label{sec::CARMA_Array_Measurements}
Normal observations with CARMA were made in parallel with the VLBI observations, and these data were used to measure the polarizations of all sources.  In full polarization mode, the CARMA correlator provides 4 GHz bandwidth (4 bands $\times$ 500 MHz $\times$ 2 sidebands) for each of the 4 polarization products (RR, LL, RL, LR) on each of the 105 baselines that connect the 15 10.4-m and 6.1-m telescopes.

The CARMA receivers operate in double sideband mode, so that signals both above and below the local oscillator frequency are downconverted to the intermediate frequency.  A 90$^\circ$ phase-switching pattern applied to the local oscillators allows these signals to be separated at the correlator.  This phase switching pattern is removed by the VLBI beamformer that sums together signals from the 8 antennas that are phased together; thus, normal observations are possible with these telescopes during VLBI scans.  However, phase switching must be disabled to the single CARMA reference antenna so that it may be treated as a completely independent VLBI station; normal CARMA observations with this antenna are possible only between VLBI scans.

Polarization observations require two additional calibrations beyond the usual gain, passband, and flux calibrations.  The first of these calibrations corrects for the delay difference between the R and L channels.  Because of the cabling in the correlator room, this delay difference varies from correlator band to correlator band as well as from antenna to antenna.  To calibrate the R-L delay, one must observe a linearly polarized source with a known position angle. Linearly polarized noise sources in the telescope receiver cabins are used for this purpose at CARMA.  Each noise source consists simply of a wire grid that is rotated into the beam in front of the receiver.  With the grid in place, horizontally polarized radiation entering the receiver originates from an ambient temperature absorber, while vertically polarized radiation originates from the sky.  Thus, there is a strong net horizontal polarization.  The L-R phase difference is measured on a channel-by-channel basis from LR autocorrelation spectra 
obtained with 
the grid in place.  Upper and lower sideband signals cannot be separated in autocorrelation spectra, but it is reasonable to assume that the R-L delays are nearly identical for corresponding channels in the upper and lower sidebands because the delays are caused almost entirely by cable length differences, which affect these signals equally.

The second polarization-related calibration is for the leakage terms of each telescope. The leakages (two complex numbers per antenna) are derived with the  {\sc MIRIAD} \cite{Sault_1995} program {\tt gpcal} from observations of a bright calibrator (polarized or unpolarized) that is observed over a wide range of parallactic angles.  We derived the leakages separately for each of the 8 correlator windows (4 bands $\times$ 2 sidebands). The VLBI data provide a rich source of leakage solutions because a number of strong sources were observed over wide parallactic angle ranges, and because the same tuning and correlator setup was used day after day.  The average leakage amplitude is 0.06, with a root mean square uncertainty of ${\pm}0.005$, where the uncertainty is estimated from the variance in 14 different leakage solutions derived over the 5 days.  The scatter in the solutions from day to day is no greater than the scatter from calibration source to calibration source on a single 
day.

For the VLBI sources, the accuracies of the polarization fraction and direction are limited by systematic errors in the leakage and R-L delay calibrations rather than by thermal noise; hence, fractional polarizations were not corrected for noise bias.  To estimate the effects of uncertainties in the leakage calibrations, we applied the ensemble of leakage solutions to the data and computed the variance in the resulting source polarizations and position angles.  Errors in the R-L delay calibration affect only the absolute EVPA of the sources.  The uncertainty in the EVPA is estimated to be ${\pm}3^{\circ}$, based on polarization measurements of Mars obtained at CARMA in 2014 May \cite{Hull_2015}.  Near the limb of the planet, emission from Mars is radially polarized \cite{Perley_Butler_2013}; deviations from radial polarization were used to estimate the absolute EVPA uncertainty.

CARMA leakage corrections for VLBI calibration were derived separately for the 2 correlator windows that were used for the ``high'' and ``low'' bands.  For the phased set of 8 antennas, the vector average of the leakages for these 8 telescopes was used.

\paragraph*{Assumptions and Procedure for Deriving a Calibration Solution.} 
To derive our complete calibration solution for the VLBI array, we made several assumptions about the data. First, we assumed that visibilities on nearly identical baselines (i.e., to the SMA/JCMT or to the phased CARMA/reference CARMA) should match. We also assumed that the fractional polarization seen on the Phased CARMA$-$Reference CARMA baseline was identical to the fractional polarization determined by CARMA for each scan. We assumed that the station gain ratios $G_{\rm R}/G_{\rm L}$ were constant over each day (with the exception of the stochastic phase drift at the SMT), and that the leakage terms at each site were constant throughout the observation (except for the modified array membership in the CARMA phased array).

To account for the unknown source polarization structure in a general way, we introduced a piecewise-linear approximation for the polarization dependence on baseline: on each baseline, the source polarization was allowed to vary linearly between points (``knots'') separated by a fixed amount of time (we used 1-hour spacings).  
We made no assumptions about the relationship between the source polarization on different baselines or the consistency of the source polarization on different days. 

After enforcing these assumptions, we performed a weighted least-squares fit to the low-band data using the linearized approximation of Eq.~\ref{eq::PolarimetricEquations}. After this fit, we dropped all data points with ${>}4\sigma$ departure from the fitted model -- usually a few percent of the data and almost exclusively very-high-SNR ($\gsim$100) quotients of parallel-hand visibilities. We then repeated the least-squares fit, typically with negligible change from the first iteration.

The high-band data could not be independently calibrated in the same way because they lack the CARMA reference antenna and because disk failures resulted in the loss of high-band data for the RCP of the CARMA phased array on days 85 and 86. Instead, we derived a high-band calibration solution by fitting the high-band data to the calibrated low-band data. As a result, the two bands can be expected to have independent thermal noise but equivalent systematic errors in the calibration solution.

\paragraph*{Estimation of Calibration Uncertainties.} 
The two major sources of uncertainty in our calibration solution are from thermal noise in the data and from the unknown source structure. We refer to these as thermal uncertainty and systematic uncertainty, respectively. The thermal uncertainty can be estimated by analyzing the $\chi^2$ hypersurface of the fitted calibration solution; the systematic uncertainty can be estimated by shifting the locations of the knots in the piecewise-linear source polarization model.

\begin{figure}[t]
\centering
\includegraphics*[width=1.0\textwidth]{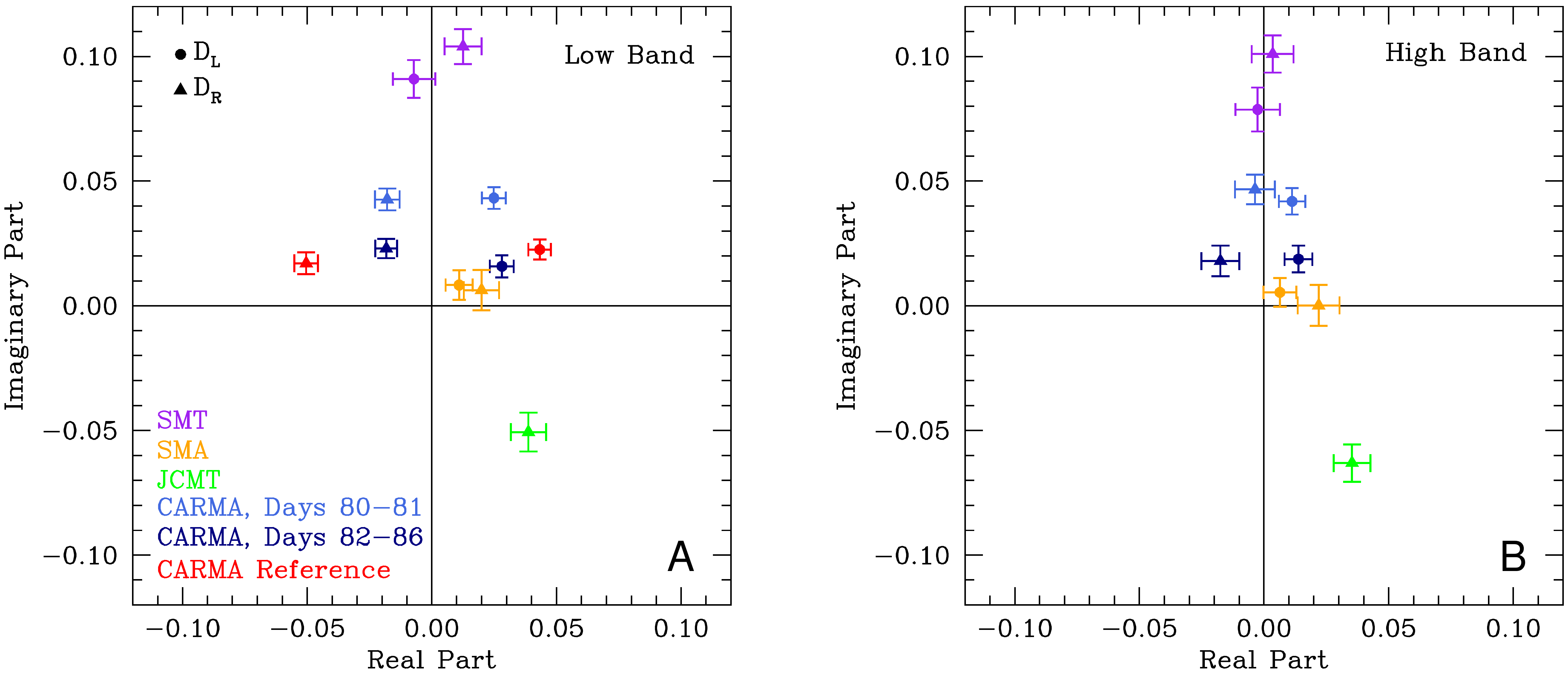}
\caption{ \small 
{\bf Polarization Leakage Terms.}  
Inferred leakage terms at all sites in our low ({\bf A}) and high ({\bf B}) observing bands. Plotted errors ($\pm1\sigma$) correspond to the cumulative thermal and structural uncertainties as determined by our Monte Carlo analysis. Solutions in the two bands show good agreement, and differences of up to a few percent are expected because of frequency structure in the leakage terms. The near symmetries in the inferred leakage pairs $\{ D_{\rm R},\, D_{\rm L}\}$ at each site are expected from general arguments, as discussed in the text.
}
\label{fig::DTerms}
\end{figure}

To determine the cumulative uncertainty, we employed a Monte Carlo error analysis. First, we successively shifted the piecewise-linear knots in increments of five minutes. Then, for each shift, we generated ten data sets in which random thermal noise equal to that of the original measurements was added to the measurements. For each such possibility, we independently derived a calibration solution. We used the mean and standard deviation of the resulting set of solutions to define our final calibration solution and uncertainty. Fig.~\ref{fig::DTerms} shows the results of our calibration analysis including uncertainties estimated from this Monte Carlo analysis.

In addition, we calibrated our data using each of these 120 derived solutions. This procedure allowed us to assess the effects of uncertainties in the calibration on the calibrated data and was important because many calibration parameters have highly covariant errors (especially $D_{\rm R}$ and $D_{\rm L}$ at different sites). 

\section*{Calibration Solution and Verification}
\label{sec::DTerms}

\begin{figure}[t]
\centering
\includegraphics*[width=1.0\textwidth]{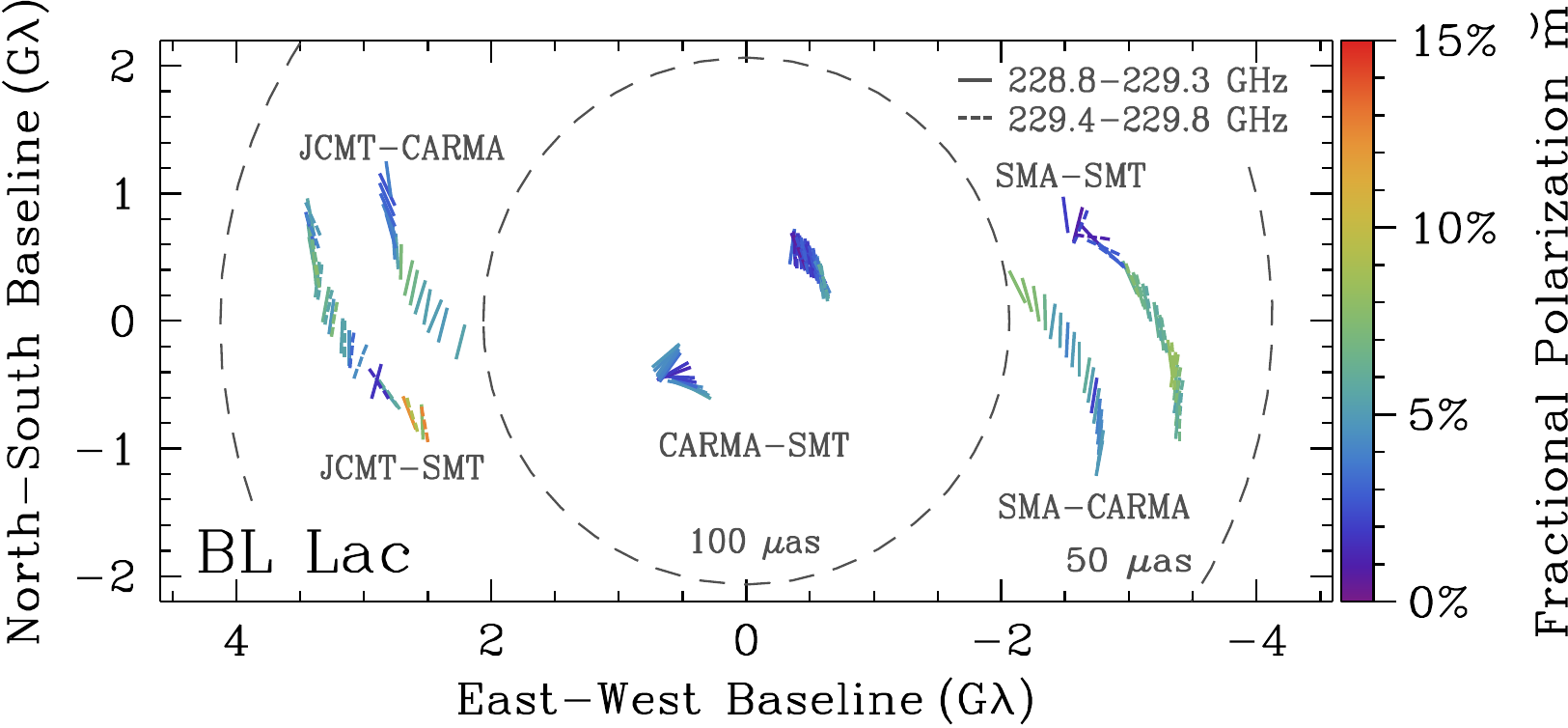}
\caption{ \small 
{\bf Calibrated interferometric fractional polarization measurements for BL~Lac.} 
BL~Lac is our most effective calibrator and shows modest (${\lsim}10\%$), slowly-varying polarization on all baselines. For each tick mark, the color  indicates the noise-debiased polarization fraction, and the angle east of north indicates the polarization direction. Because of its low polarization, BL~Lac is especially sensitive to calibration errors. For maximal clarity, we omit baselines to the CARMA reference antenna and measurements on short (30-second) scans. The low polarization on all baselines reinforces the appropriateness of the linear approximation of Eq.~\ref{eq::PolarimetricEquations}.
}
\label{fig::BLLAC_Pol}
\end{figure}

We now discuss our polarimetric calibration solution, including comparisons with independent leakage estimates by our participating arrays and the apparent symmetries. Fig.~\ref{fig::BLLAC_Pol} shows the calibrated data for BL Lac, which was the most important source for deriving the calibration solution.

\paragraph*{Comparison with Independent Leakage Estimates.} 
\label{sec::DTerms_Compare}

The phased arrays, CARMA and the SMA, can independently estimate their leakage terms. However, for both of these arrays, each element has identical field rotation, leading to a degeneracy in the polarimetric equations (Eq.~\ref{eq::PolarimetricEquations}): adding an offset $\Delta$ to every $D_{i, \rm R}$ in the array and subtracting $\Delta^\ast$ from every $D_{i,\rm L}$ will preserve these equations. Because EHT baselines join stations with different field rotations, they break this degeneracy. Nevertheless, suitable combinations of leakage terms, most simply $D_{\rm R} + D_{\rm L}^\ast$, still provide meaningful comparisons between our calibration solution and those derived by the arrays. 

At CARMA, the leakage terms show structure, or ``ripples'', across the passband. We therefore compare EHT calibration solutions with the average of CARMA-determined values across the appropriate frequency range. EHT calibration results for the sum of the leakage terms are consistent with the CARMA results to ${\sim}$0.01, and to within 0.002 on later days, which include the long scans on BL Lac (see Table \ref{tab::CARMA_Dterms}). Moreover, the inferred polarization on the Phased CARMA-Reference CARMA baseline matches the values determined by CARMA, even for sources (such as \sgra) that were not included in the calibration.

At the SMA, the leakage terms have a small (${\sim}2\%$) random component plus a linear slope in their real part with frequency, consistent with the expected leakage incurred by departures from the tuned frequency of the QWPs \cite{Marrone_Thesis}. The imaginary parts of the leakage terms may reflect orientation errors in the QWPs.  Best estimates using a combination of historical data and other calibration measurements from 2013 give $D_{\rm L} \approx 0.015-0.009i - \Delta^\ast$ and $D_{\rm R} \approx 0.018+0.005i + \Delta$ for some unknown $\Delta$; however, the imaginary parts of these measurements are uncertain to ${\sim}1\%$ because of slight orientation shifts of the QWPs upon successive re-installations. The average real part is then 0.017, which is quite close to the average of 0.015 from the EHT calibration solution. The difference in imaginary parts differs from the EHT calibration solution by $2\%$, consistent with the effects of different QWP installations.

This independent analysis for CARMA and the SMA was done in {\sc MIRIAD} \cite{Sault_1995}, which follows slightly different conventions than ours. To fix the complex degree of freedom in connected-element calibration, {\sc MIRIAD} imposes the constraint $\sum_i D_{i,\rm R} = \sum_i D_{i,\rm L}^\ast$. Also, the field rotation angles used in {\sc MIRIAD} differ by a constant offset $\delta \phi$ from our conventions, rotating $D_{\rm R}$ by $2 \delta\phi$ and $D_{\rm L}$ by $-2\delta \phi$ ($\delta \phi$ is specified via the {\sc MIRIAD} header parameter \texttt{evector}). Lastly, {\sc MIRIAD} employs an opposite baseline ordering convention to ours, introducing an overall conjugation of both $D_{\rm R}$ and $D_{\rm L}$ \cite{Sault_1994}.

\begin{table}[t]
\centering
{
\begin{tabular}{lcc}
\toprule
\textbf{Site} & CARMA $D_{\rm R}+D_{\rm L}^\ast$ & EHT $D_{\rm R}+D_{\rm L}^\ast$\\
\midrule
CARMA Reference Antenna           & $-0.009 + 0.000 i$ & $-0.008 -  0.006 i$\\
CARMA Phased Array, Days 080-081  & $\hphantom{-}0.013 + 0.003 i$ & $\hphantom{-}0.001 -  0.001 i$\\
CARMA Phased Array, Days 082-086  & $\hphantom{-}0.010 + 0.006 i$ & $\hphantom{-}0.010 +  0.007 i$\\
\bottomrule
\end{tabular}
}
\caption{Comparisons of our low-band leakage solution for CARMA with independent estimates using only CARMA data.}
\label{tab::CARMA_Dterms}
\end{table}

\paragraph*{Symmetries in the Leakage Terms.}\label{sec::Symmetries} 
Our derived calibration solution exhibits clear symmetries between the left and right leakage terms. Because such symmetries are not included as prior constraints in our calibration procedure but are expected from general considerations, they provide additional validation for the calibration solution. 

For example, the inferred leakage terms at CARMA and the SMT have $D_{\rm R} \approx -D_{\rm L}^\ast$. This symmetry is expected whenever the Jones matrix acts in such a way that the power is preserved and the outputs are orthogonal (i.e., if $\textbf{J}$ is unitary: $\textbf{J}^{-1} = \textbf{J}^\dag$) (see Appendix D of \cite{RWB94} or \cite{HagforsCampbell74}). Indeed, many instrumental effects produce a unitary Jones matrix. Common examples include any rotation of orthogonal feeds or of a QWP, and departures from the tuned frequency of a QWP.

For sites that do not have orthogonal feeds (i.e., the SMA and the JCMT), the situation is slightly different. For instance, the general action of the QWP system at the SMA is described in \cite{Marrone_Thesis} and gives leakage terms that satisfy $D_{\rm R} = D_{\rm L}^\ast$ (when an error in the QWP retardation is included but not in the field transmission), regardless of the orientation of the feed or of the QWP. As a specific example of this relationship, for errors that only arise from a frequency offset from the tuned frequency of the QWP, the left and right leakage terms will be real and equal. Note that this relationship differs from the previous result ($D_{\rm R} = -D_{\rm L}^\ast$) because of the feed {configuration} rather than the feed (or QWP) {orientation}.

\paragraph*{Possible Elevation Dependence of Leakage Terms.}\label{sec::Elevation_Dependence} 
Our current calibration solution has no allowance for elevation dependence of the leakage terms. Nevertheless, several features in the calibrated data suggest that elevation-dependent leakage effects are small. Most directly, the close agreement of the fractional polarization on the SMT-CARMA baseline with CARMA-only measurements of \sgra\  (e.g., Fig.~\ref{fig::SMTCARMA_CARMA_Compare}) shows that neither the SMT nor CARMA can have large elevation dependence for their leakage terms. This result is especially secure because the \sgra\ data was not used to derive the calibration solution. Elevation-dependent leakage at the SMA and JCMT would appear in the comparison between redundant measurements to the two stations (when the SMA sampled RCP) and also in non-simultaneous amplitude ratios on the JCMT-SMA baseline (discussed above). These comparisons constrain elevation-dependent leakage changes to be ${\lsim}$0.01 for our current data at all sites -- 
comparable to the derived uncertainties of our leakage terms (Fig.~\ref{fig::DTerms}) and of negligible importance for our current results.

\paragraph*{Additional Calibration Considerations for Sgr A*.} 
Although the low polarization of BL~Lac justifies the linear approximation for the polarimetric equations (Eq.~\ref{eq::PolarimetricEquations}) when deriving our calibration solution, the higher polarization fractions that we observe on long baselines to \sgra\ can introduce errors from higher-order terms. Nevertheless, higher order corrections to the inferred $\breve{m}(\textbf{u})$ are proportional to the products of $\breve{m}(\textbf{u})^2$ and $\breve{m}(\textbf{u}) \breve{m}(-\textbf{u})$ with a leakage term, and these corrections are within our current thermal noise. 

Another consideration for the long-baseline measurements of \sgra\ is the effects of circular polarization. To leading order, circular polarization scales the parallel-hand visibilities by $\left( 1 \pm \breve{v} \right)$, where $\breve{v} \equiv \tilde{\mathcal{V}}/\tilde{\mathcal{I}}$ is the fractional circular polarization in the visibility domain and +/- corresponds to RCP/LCP. 
Although the fractional image-averaged circular polarization for \sgra\ at $\lambda$=1.3~mm is only ${\approx}$1\% \cite{Munoz_2012}, the fractional circular polarization $\breve{v}$ on long baselines could be an order of magnitude higher, just as we observe for the fractional linear polarization $\breve{m}$. 

We can robustly estimate $\breve{v}$ by comparing pairs of RCP and LCP parallel-hand visibilities on baselines to the SMA, using adjacent long/short scans with the SMA sampling opposite polarization states.  
However, because the short scans have low-SNR, we have only five measurements suitable for comparison. These measurements show no statistically significant detection of circular polarization, with thermal noise giving 1$\sigma$ uncertainties ranging from 11\% to 20\% for individual estimates of $\breve{v}$. Even within these crude limits, the allowed circular polarization is insufficient to produce the distinctive asymmetry that we see in $\breve{m}(\pm\textbf{u})$ on long baselines, often in excess of a factor of three difference, which would require an implausibly large $\breve{v} \gsim 50\%$. 

The circular polarization $\breve{v}$ can also be estimated using only long scans by comparing opposite parallel-hand visibilities on baselines to the SMA (LCP) with those to the JCMT (RCP). However, this comparison is dominated by relative uncertainties in the time-variable gains at these sites (which do not affect $\breve{m}$).

\section*{Amplitude Calibration Procedure and Results.} 

Our amplitude calibration (i.e., the estimation of $|\tilde{\mathcal{I}}(\textbf{u})|$) followed a similar methodology to previous work with the EHT \cite{Fish_2011,Doeleman_2012,Lu_2013} and will be presented in detail elsewhere. We now briefly summarize the calibration procedure and results.

\paragraph*{Amplitude Calibration Strategy.} After correlation, we estimated parallel-hand visibility amplitudes via incoherent averaging \cite{RDM95}. Incoherent averaging begins by coherently averaging visibilities over a specified segmentation time and then incoherently averages the resulting set of visibilities. To avoid a loss of signal, the segmentation time must be shorter than the timescale of differential phase fluctuations between sites; in our case, the dominant fluctuations are from the atmosphere, with a typical timescale of 5-10 seconds. In addition, while incoherent averaging provides a nearly optimal estimator of the signal amplitude when the SNR of each averaged segment is high, it is severely sub-optimal when the segment SNR is low. Because of these two considerations, we first performed incoherent averaging with a 1-second segmentation time; for scans with a resulting SNR less than 15, we then repeated the incoherent averaging with a segmentation time of 6 seconds. 

We next converted the estimated visibilities to physical units of correlated flux density by applying an a~priori calibration solution. This solution required estimates of the antenna gain (Jy/K), system temperature, and atmospheric opacity for each station and each scan. For the participating phased arrays (CARMA and the SMA), we also included scan-by-scan estimates of the phasing efficiency. Because our present analysis is focused on the normalized visibility -- i.e., the correlated flux density on a given baseline divided by the simultaneous correlated flux density on a zero-baseline -- the conversion to physical units does not affect our results but time variability of the gains does. 
 
To solve for the remaining time-variable gains at each station and in each circular polarization, we assumed that the gains have a constant ratio $|G_{\rm R}|/|G_{\rm L}|$ per station per day -- an identical assumption to our polarization calibration and an assumption that was validated by comparing the RCP and LCP parallel-hand correlations between dual-polarization stations. Next, we merged consecutive short/long scans. Doing so provides an effective parallel-hand, zero-baseline interferometer on Hawaii (SMA-JCMT) during the short scans that can be used to estimate the gains for the adjacent long scans. For each merged scan, we then solved for a calibration solution using a least-squares minimization of all measurements. For these solutions, we assumed that the circular polarization is zero, although our resultant solution would not eliminate the imprint of circular polarization because we do not solve for separate RCP and LCP gains.

The resulting gain solution is quite robust, a consequence of the particular baseline redundancy of the EHT. To understand the reason for this robustness, consider two pairs of co-located stations: $\{A,A'\}$ and $\{B,B'\}$. [By co-located, we simply mean that the visibility $V_{A A'}$ is equal to what would be measured by a zero-baseline interferometer.] These provide two independent closure amplitudes:
\begin{align}
\mathcal{A}_1 &= \left| \frac{ V_{A B} V_{A' B'}}{ V_{A B'} V_{A' B}} \right|\\
\nonumber \mathcal{A}_2 &= \left| \frac{ V_{A B} V_{A' B'}}{ V_{A A'} V_{B B'}} \right|,
\end{align}
where $V_{A B}$ denotes a visibility on the baseline joining stations $A$ and $B$. Closure amplitudes such as these are especially valuable because they are immune to station-based gain fluctuations \cite{TMS}. The first of these is a ``trivial'' closure amplitude: $\mathcal{A}_1=1$. However, the second, $\mathcal{A}_2 = \left| V_{AB}/V_{A A'} \right|^2$, gives the squared normalized visibility on the $A$--$B$ baseline. As a result, the normalized visibility on any baseline joining two pairs of co-located sites is a closure quantity. The limiting noise on such baselines will likely be thermal noise rather than systematic uncertainties. For our current data, the phased CARMA/reference CARMA and SMA/JCMT pairs provide this degree of redundancy, so we can estimate the CARMA-SMA/JCMT normalized visibility as a closure quantity.

\begin{figure}[t]
\centering
\includegraphics*[width=0.6\textwidth]{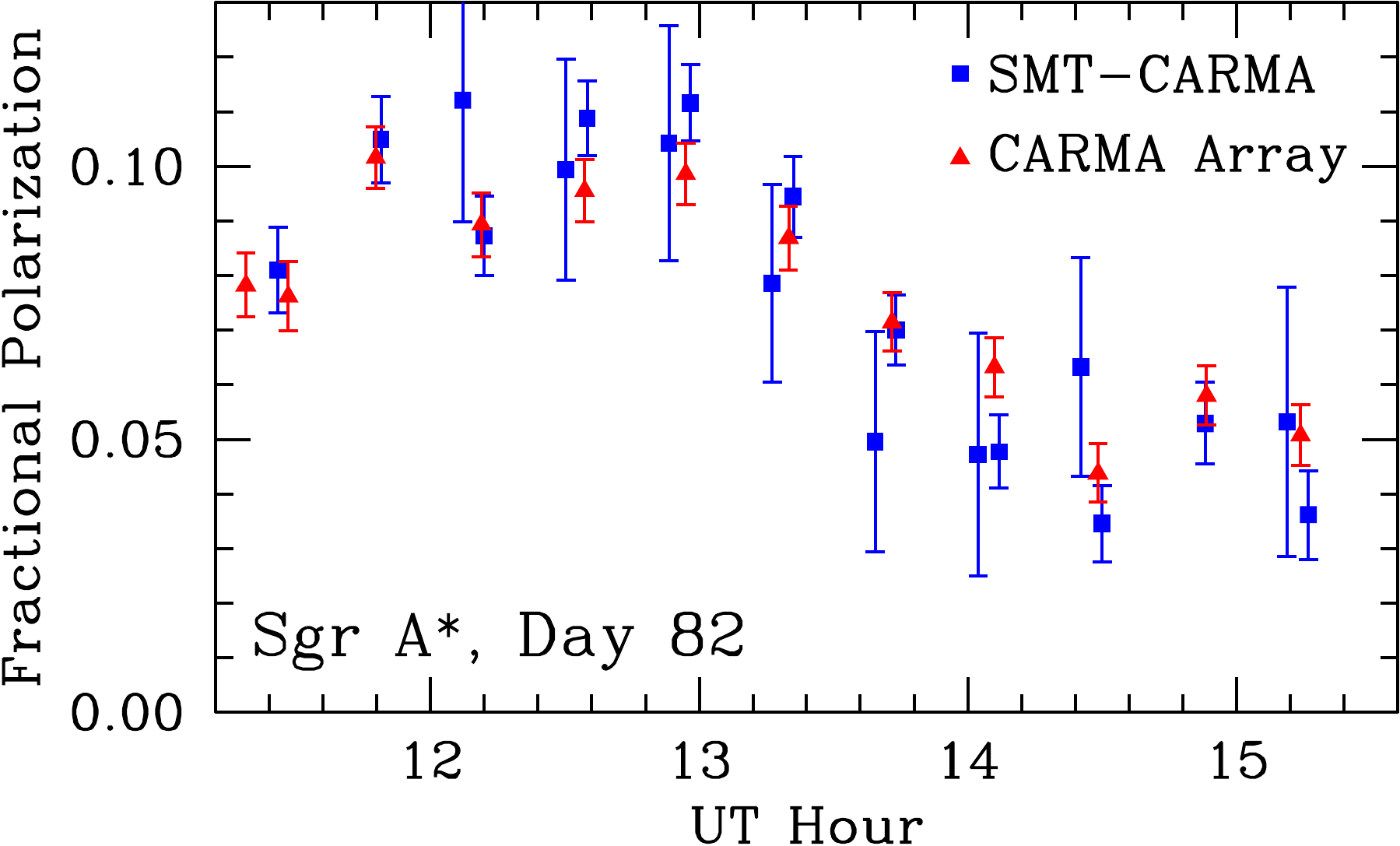}
\caption{\small 
{\bf Comparison of CARMA-only and SMT-CARMA fractional polarization measurements for \sgra.} 
Comparison of the calibrated fractional polarization measurements made by CARMA with those measured on the SMT-CARMA baseline for \sgra\ on one day. For the SMT-CARMA baseline, we averaged the pair of measurements $\breve{m}(\pm \textbf{u})$ to eliminate the linear dependence of $\breve{m}$ on baseline (see Eq.~\ref{eq::m_leading}). 
Errors ($1\sigma$) denote the aggregate of thermal and systematic uncertainties; systematic uncertainties ($0.5\%$) are dominant for measurements by CARMA but are subdominant for those of the SMT-CARMA baseline. Fitting all measurements for a single relative factor $f$ in the polarizations ($f \equiv \breve{m}_{\rm CARMA}/\breve{m}_{\rm SMT-CARMA}$) gives $f = 0.97 \pm 0.03$ on this day, with a reduced chi-squared $\chi^2_{\rm red} \approx 0.79$ ($\chi^2_{\rm red} < 1$ may suggest that systematic errors for CARMA are overestimated). This close agreement, $f \approx 1$, rules out the possibility of substantial unpolarized or differentially polarized emission (relative to the compact emission) on scales between a few hundred $\mu$as and ${\sim}10''$. Such emission could not contribute more than ${\sim}10\%$ of the total flux in these observations.
}
\label{fig::SMTCARMA_CARMA_Compare}
\end{figure}

However, no visibility on a baseline to the SMT can be considered known or can be compared to other identical measurements. As a result, the gain calibration of the SMT cannot be improved beyond the a~priori calibration solution without additional assumptions about source structure. Nevertheless, our polarization measurements suggest one robust pathway to determine an approximate gain solution at the SMT (to within ${\sim}$10-20\%). 
Specifically, our polarization measurements argue that \sgra\ is so compact that we can approximate the SMT-CARMA baseline as yet another zero-baseline (i.e., a baseline that does not resolve the image) (see Fig.~\ref{fig::SMTCARMA_CARMA_Compare}).   
Because the visibility amplitude $|\tilde{I}(\textbf{u})|$ is always maximum when $\textbf{u}=\textbf{0}$, this strategy determines a strict upper limit for visibility amplitudes on baselines to the SMT. Thus, to conservatively bracket the remaining uncertainty, we derive our gain solution by assuming that the SMT-CARMA visibility amplitude is 90\% of the zero-baseline value, and we add 10\% systematic uncertainty at quadrature to all baselines to the SMT. [While we derive a time-variable SMT gain solution by approximating SMT-CARMA as a zero-baseline interferometer, we do not assume that the SMA/JCMT-CARMA and SMA/JCMT-SMT baselines should yield identical visibilities.]

With this final approximation, we can independently derive a gain solution for any merged scan on \sgra\ that includes detections on the SMA-JCMT baseline, the Phased CARMA-Reference CARMA baseline, the SMT-Phased CARMA baseline, the SMT-Reference CARMA baseline, and a pair of duplicate baselines to the SMA and the JCMT from either CARMA or the SMT. When the visibilities are normalized to the zero-baseline flux, the a priori station gains and absolute zero-baseline flux estimate are irrelevant because the calibrated normalized visibilities are now solved for self-consistently. 

As noted earlier, fractional circular polarization will bias the parallel-hand visibilities by a factor of $\left(1\pm \breve{v}\left(\textbf{u}\right) \right)$ on each baseline $\textbf{u}$. Linear polarization will bias the visibilities by a factor of, e.g., $\left(1 + \breve{m}(\textbf{u}) D_{1,{\rm L}} e^{-2i \phi_1} + \breve{m}^\ast(-\textbf{u}) D_{2,{\rm L}}^\ast e^{2i \phi_2}\right)$. To account for remaining systematic uncertainties from these polarization contributions, we added 10\% of each estimated visibility amplitude $|\tilde{\mathcal{I}}|$ at quadrature to its thermal noise. 

\paragraph*{Scattering Mitigation Procedure.} To meaningfully compare values of fractional polarization, $\breve{m}$, and correlated total flux, $\tilde{\mathcal{I}}$, we must first account for scatter broadening (or ``blurring'') of the image (see \cite{Fish_2014} for details). Mathematically, this scatter broadening convolves the unscattered images of each Stokes parameter with a kernel $G(\textbf{x})$; equivalently, scatter broadening multiplies the unscattered source visibilities by the Fourier conjugate scattering kernel,  $\tilde{G}(\textbf{u}) {\leq} 1$. Thus, scatter broadening introduces a baseline-dependent decrease in correlated flux. If the scattering kernel is known, then these effects of scattering can be inverted by multiplying measured visibilities by a ``deblurring'' factor $1/\tilde{G}(\textbf{u})$ \cite{Fish_2014}. The form of $G(\textbf{x})$ (and, hence, of $\tilde{G}(\textbf{u})$) is often assumed to be an elliptical Gaussian. 

Because scatter broadening affects all polarizations equally, scatter broadening does not affect $\breve{m}(\textbf{u})$, but it does affect $\tilde{\mathcal{I}}(\textbf{u})$. To invert the effects of scattering on our measured values of $\tilde{\mathcal{I}}(\textbf{u})$, we used the elliptical Gaussian scattering kernel $\tilde{G}(\textbf{u})$ from \cite{Bower_2006}: a FWHM of $1.309 \lambda_{\rm cm}^2~{\rm mas}$ along the major axis, at a position angle of $78^\circ$ (east of north), and a FWHM of $0.64\lambda_{\rm cm}^2~{\rm mas}$ along the minor axis. For the SMT-CARMA baseline, $1/\tilde{G}(\textbf{u}) \leq 1.02$, so the effects of scattering are within our calibration uncertainties. For baselines from the SMT and CARMA to the SMA and JCMT, $1/G(\textbf{u})$ ranges from 1.35 (on the shortest baselines) to 1.67 (on the longest baselines).

Uncertainties in the size of the scattering ellipse at longer wavelengths have little effect on these deblurring factors (all within our stated systematic uncertainties), especially because our long baselines are predominantly east-west, where the scattering properties are precisely known. However, the extrapolation of the scattering law to $\lambda{=}1.3~{\rm mm}$ carries larger systematic uncertainties. 
For example, measurements of scattering-induced substructure in the image of \sgra\ at $\lambda=1.3~{\rm cm}$ suggest that the scattering transitions from a square-law regime to a Kolmogorov regime at wavelengths shorter than $\lambda=1~{\rm cm}$ \cite{Gwinn_2014,Johnson_Gwinn_2015}. This transition will affect the scaling of scatter-broadening with frequency and also the dependence of the kernel on baseline. The former causes the angular broadening to scale as $\lambda^{11/5}$ rather than $\lambda^{2}$; the latter causes $\tilde{G}(\textbf{u})$ to fall as $e^{-|\textbf{u}|^{5/3}}$ rather than as $e^{-|\textbf{u}|^{2}}$. 
Each of 
these effects then implies that the scattering at $\lambda{=}$1.3~mm may be weaker than we have assumed. As a result, the long-baseline visibilities $|\tilde{\mathcal{I}}|$ of the unscattered image may be systematically lower than we have estimated, by $10-25\%$ depending on baseline length. However, this systematic uncertainty has little effect on the comparison between $|\tilde{\mathcal{I}}(\textbf{u})|/\tilde{G}(\textbf{u})$ and $\breve{m}(\textbf{u})$ that we use to estimate the degree of order in the polarization vector field (Fig.~\ref{fig::flux_fpol}).

\begin{figure}[t]
\centering
\includegraphics*[width=0.7\textwidth]{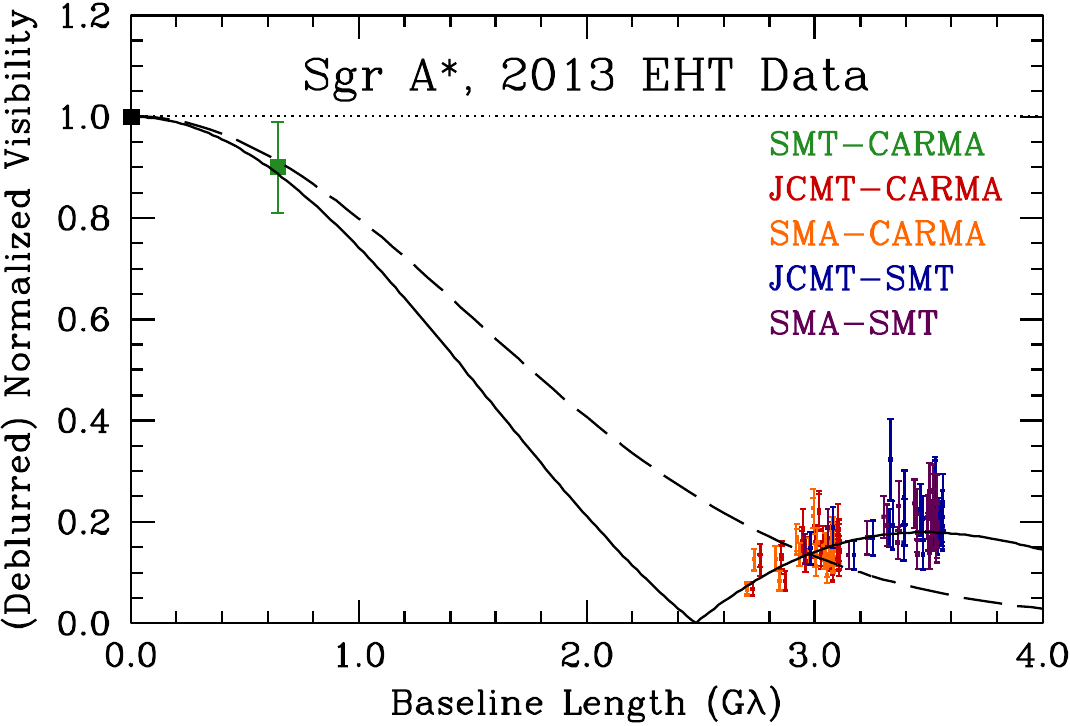}
\caption{ \small 
{\bf Total-intensity VLBI of \sgra\ and fitted geometrical models.} Normalized, deblurred visibilities are shown as a function of baseline length; errors are $\pm 1\sigma$ (thermal plus systematic). The dashed line shows the best-fit circular Gaussian (FWHM: 52~$\mu$as), which is incompatible with our data. The best-fit elliptical Gaussian provides a fit that is only marginally better. Many simple geometrical models can produce a rising visibility with baseline at ${\sim}$2.7~G$\lambda$. An annulus of uniform intensity (inner diameter: 21~$\mu$as, outer diameter: 97~$\mu$as), shown with a solid line, is perhaps the simplest model that is consistent with our data, although many alternatives are possible (see discussion in the text). For most points, remaining systematic uncertainties in the calibration solution are the dominant source of error. 
}
\label{fig::amplitudes}
\end{figure}

\paragraph*{Calibrated Visibilities for \sgra.} 
Fig.~\ref{fig::amplitudes} shows our calibrated visibility amplitudes; a comprehensive analysis of these amplitudes will be presented elsewhere. The improved sensitivity of our experiment relative to prior years \cite{Doeleman_2008,Fish_2011} provides measurements at lower source elevation on baselines to Hawaii, significantly extending the range of baseline lengths sampled. The extended coverage shows that the deblurred amplitudes do not fall monotonically with increasing baseline length and are inconsistent with a single Gaussian component. The rise in the visibility amplitude with baseline at ${\sim}$2.7~G$\lambda$ determines an absolute minimum of 38~$\mu$as for the total east-west extent of the flux. Although the measurements near 2.7~G$\lambda$ occur when \sgra\ is at low elevation for CARMA and must be interpreted with appropriate caution, those lowest visibilities are coincident with the highest fractional polarizations (cf.~Fig.~\ref{fig::CARMA_Long_Compare}). The polarization measurements thereby 
support the visibility minimum at ${\sim}$2.7~G$\lambda$ without requiring assumptions about the scattering kernel or the gain calibration.

However, the overall emission structure is not yet uniquely determined because of the sparse and exclusively east-west long-baseline coverage, and many disparate models provide a reasonable fit to the data. [By the projection-slice theorem, a baseline only samples structure projected along its direction.] One such example is a circular Gaussian (FWHM: 54~$\mu$as) plus a point source with 24\% of the Gaussian's integrated flux and offset 43~$\mu$as to the east or west. Another is two uniform-intensity disks centered on the origin with diameters of $12~\mu{\rm as}$ and $155~\mu{\rm as}$ and contributing 14\% and 86\% of the total flux, respectively. We therefore focus on deriving general conclusions about the polarization structure and order (e.g., Fig.~\ref{fig::flux_fpol}) that do not assume a specific model for the intensity.

\section*{Additional Discussion of Geometric Constraints}

We now provide additional details on the relationship between fractional polarization in the visibility and image domains and derive the geometrical constraints discussed in the main text. As these cases demonstrate, even sparse fractional polarization information in the visibility domain can provide robust insights into the corresponding fractional polarization in the image domain. Fig.~\ref{fig::dictionary} shows some characteristic examples.

\afterpage{
\begin{figure}[th]
\centering
\includegraphics*[width=0.6\textwidth]{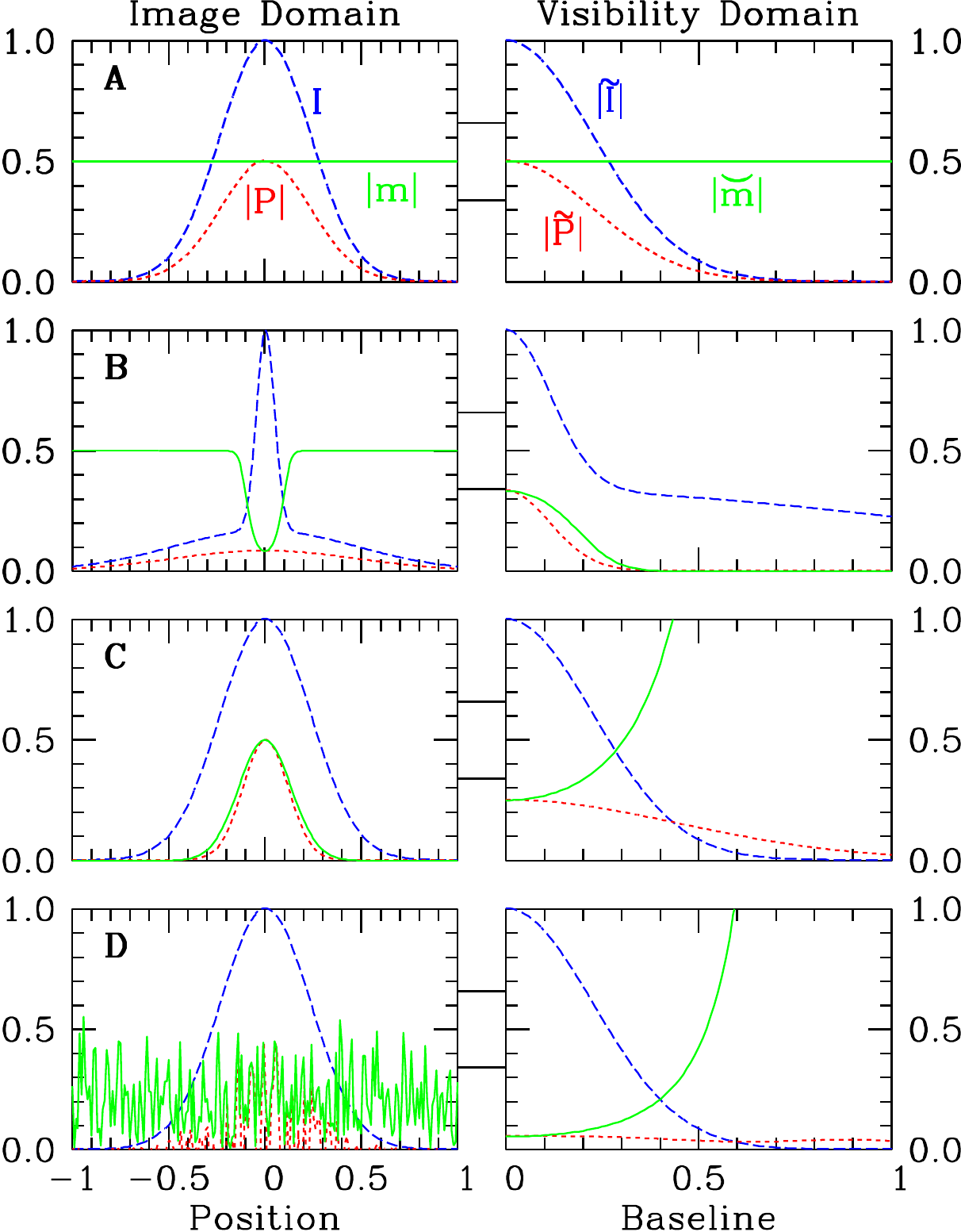}
\caption{\small 
{\bf Examples of the fractional polarization in the image and visibility domains.} 
Total intensity ($I$), polarization ($P$), and fractional polarization ($m$) in the image and visibility domains corresponding to four models, each with a Gaussian distribution of total intensity. {\bf A.)} Constant fractional polarization of $50\%$: The fractional polarization is constant and equal in both the image and the visibility domains. {\bf B.)} Unpolarized core with a diffuse halo of constant fractional polarization: Long baselines resolve the diffuse component, so the fractional polarization in the visibility domain asymptotes to the fractional polarization of the most compact component -- in this case, zero. {\bf C.)} Gaussian intensity distribution with a tapered polarized component: In this example, the taper of the polarized component allows the signal to persist on longer baselines 
than for the more smoothly varying total intensity, causing the fractional polarization in the visibility domain to rise monotonically to infinity. {\bf D.)} Disordered polarization throughout a Gaussian image: The Gaussian taper in the image imparts a broad correlation structure to the polarization in the visibility domain without diminishing its amplitude. Short baselines see little polarization because of cancellations, but this small value determines a noise floor that persists to long baselines. Parseval's theorem  -- the conservation of integrated power between the two domains -- then relates the relative amplitudes in the image and visibility domains to the relative extents in each domain. For example, if the zero-baseline fractional polarization is (on average) a factor of 10 weaker than the average fractional polarization of the image, then $\tilde{\mathcal{Q}}+i\tilde{\mathcal{U}}$ will persist to baselines that are 10 times longer than those that resolve $\tilde{\mathcal{I}}$. 
}
\label{fig::dictionary}
\end{figure}
\clearpage
}

As introduced in the main text and again in this supplement, fractional polarization in the visibility domain is defined as $\breve{m}(\textbf{u}) \equiv \left[\tilde{\mathcal{Q}}(\textbf{u}) + i \tilde{\mathcal{U}}(\textbf{u})\right]/\tilde{\mathcal{I}}(\textbf{u})$. Note that each of the Stokes parameters is real, so each obeys a conjugation relationship in the visibility domain (e.g., $\tilde{\mathcal{Q}}(-\textbf{u}) = \tilde{\mathcal{Q}}^\ast(\textbf{u})$). However, $\tilde{\mathcal{Q}}(-\textbf{u}) + i \tilde{\mathcal{U}}(-\textbf{u}) = \tilde{\mathcal{Q}}^\ast(\textbf{u}) + i \tilde{\mathcal{U}}^\ast(\textbf{u}) \neq \left[\tilde{\mathcal{Q}}(\textbf{u}) + i \tilde{\mathcal{U}}(\textbf{u})\right]^\ast$, so $\breve{m}(-\textbf{u})\neq \breve{m}^\ast(\textbf{u})$. Indeed, an easy way to see that reversed baselines carry distinct polarimetric information is to form the following mixed quantities:
\begin{align}
\frac{1}{2} \left[ \breve{m}(\textbf{u}) + \breve{m}^\ast(-\textbf{u}) \right] &= \frac{\tilde{\mathcal{Q}}(\textbf{u})}{\tilde{\mathcal{I}}(\textbf{u})}\\
\nonumber \frac{1}{2i} \left[ \breve{m}(\textbf{u}) - \breve{m}^\ast(-\textbf{u}) \right] &= \frac{\tilde{\mathcal{U}}(\textbf{u})}{\tilde{\mathcal{I}}(\textbf{u})}.
\end{align}
These identities also demonstrate that, given dual-polarization measurements at each station, the Stokes $\mathcal{Q}$ and $\mathcal{U}$ images can be independently reconstructed \cite{Conway_1969}.

\paragraph*{Polarization Morphology Constraints.} 
We now consider how specific polarization morphologies in an image manifest themselves in visibilities. First, suppose that the complex fractional polarization is constant across the image: $\mathcal{Q}(\textbf{x}) + i \mathcal{U}(\textbf{x}) \equiv m_0 \mathcal{I}(\textbf{x})$. In this case, $\tilde{\mathcal{Q}}(\textbf{u}) + i \tilde{\mathcal{U}}(\textbf{u}) \equiv m_0 \tilde{\mathcal{I}}(\textbf{u})$, so $\breve{m}(\textbf{u}) = m_0$. Hence, an image with constant fractional polarization will produce visibilities with the same constant fractional polarization. 

A second case of interest occurs when the fractional polarization has constant direction throughout the image. After a constant rotation of the polarization vectors, this assumption is equivalent to taking $\mathcal{U}(\textbf{x}) \equiv 0$. In this case, the polarization obeys $\breve{m}(-\textbf{u}) = \breve{m}^\ast(\textbf{u})$, since both $\tilde{\mathcal{Q}}$ and $\tilde{\mathcal{I}}$ obey this relationship. More generally, for an arbitrary polarization direction $\varphi \equiv \frac{1}{2} {\rm arg}(m_0)$, we obtain $\breve{m}(-\textbf{u}) = e^{4i \varphi} \breve{m}^\ast(\textbf{u})$. So, whenever the polarization direction is constant across an image, the fractional polarization in the visibility domain will satisfy $\left| \breve{m}(-\textbf{u}) \right| = \left| \breve{m}(\textbf{u}) \right|$.

\paragraph*{Polarization Centroid Constraints.} 
Next, we consider the geometrical constraints discussed in the main text for the relative offsets of the polarized and the total flux. A more precise and extensive discussion with a focus on time variability is given in \cite{Johnson_2014}. 

For measurements of the correlated flux $\tilde{\mathcal{I}}(\textbf{u})$ on a short baseline $\textbf{u}$, the leading-order contribution to phase comes from the image centroid, ${\left \langle \textbf{x}\, \mathcal{I}(\textbf{x}) \right \rangle}/{\left \langle \mathcal{I}(\textbf{x}) \right \rangle}$:
\begin{align}
\label{eq::Centroid}
\tilde{\mathcal{I}}(\textbf{u}) \equiv \int d^2\textbf{x}\, \mathcal{I}(\textbf{x}) e^{-2\pi i \textbf{u} \cdot \textbf{x}} &\approx \left \langle \mathcal{I}(\textbf{x}) \right \rangle\left(1 -2\pi i \textbf{u} \cdot \frac{\left \langle \textbf{x}\, \mathcal{I}(\textbf{x}) \right \rangle}{\left \langle \mathcal{I}(\textbf{x}) \right \rangle} \right),
\end{align}
where $\textbf{x}$ denotes sky coordinates, in radians. Note that an offset of the centroid by the nominal resolution ($1/|\textbf{u}|$) of an interferometric baseline $\textbf{u}$ produces a full wrap of phase. 

The same expansion holds for polarization, $\mathcal{P} \equiv \mathcal{Q} + i \mathcal{U}$. Thus, to leading order, the baseline dependence of fractional polarization can be written in the following form:
\begin{align}
\label{eq::m_leading}
\breve{m}(\textbf{u}) &\approx \breve{m}(\textbf{0}) \left[ 1 + 2\pi i \textbf{u} \cdot \left( \frac{ \left \langle \textbf{x}\, \mathcal{I}(\textbf{x}) \right \rangle }{ \left \langle \mathcal{I}(\textbf{x}) \right \rangle } - \frac{ \left \langle \textbf{x}\, \mathcal{P}(\textbf{x}) \right \rangle }{ \left \langle \mathcal{P}(\textbf{x}) \right \rangle } \right) \right].
\end{align}
Hence, the leading-order correction rotates the phase of $\breve{m}(\textbf{u})$ linearly with the offset between the centroids of the polarized and the total flux projected along $\textbf{u}$.

\begin{figure}[t]
\centering
\includegraphics*[width=0.8\textwidth]{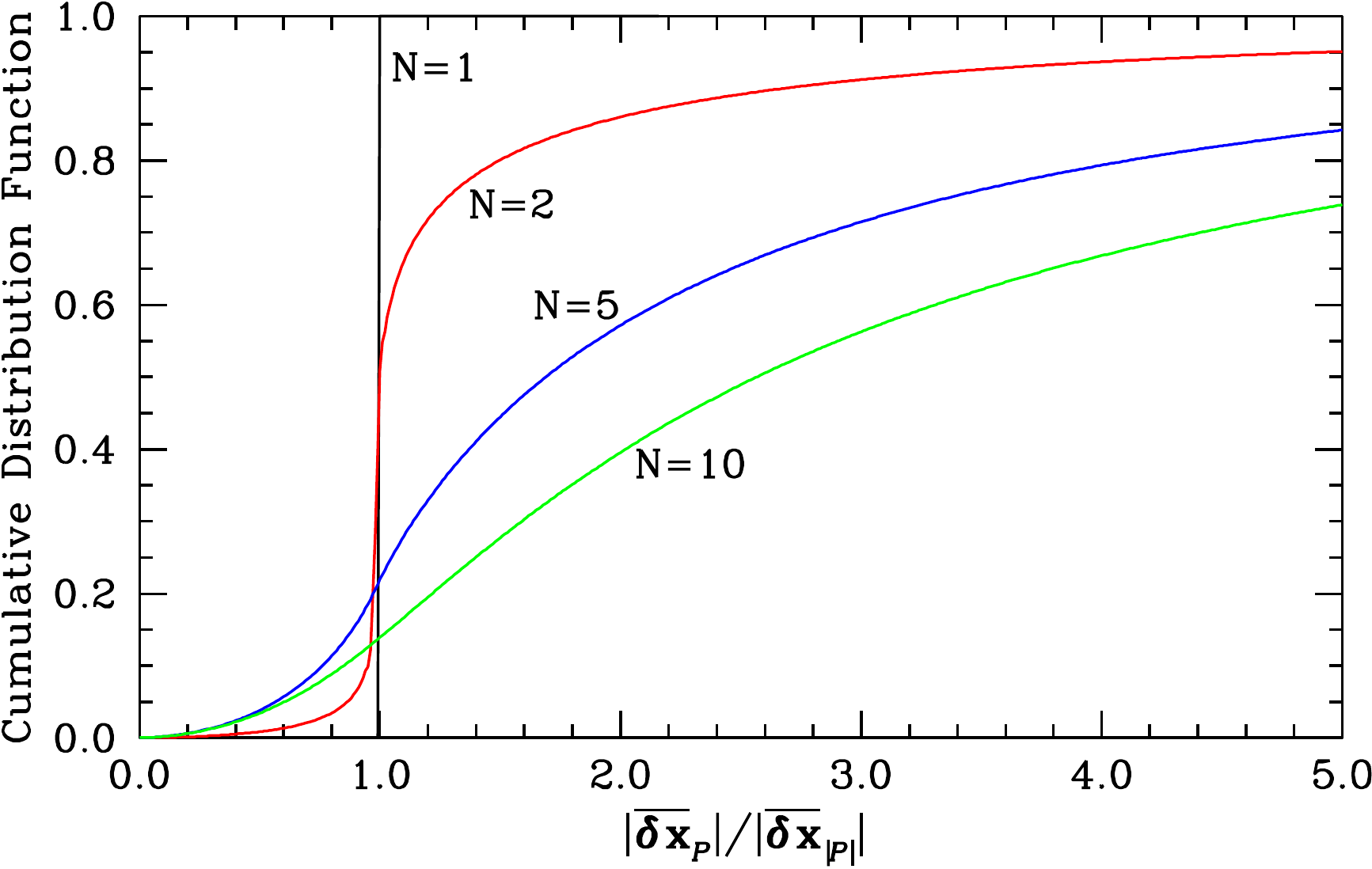}
\caption{ \small 
{\bf Relative amplitude of two polarization offset estimators.} 
Cumulative distribution function (CDF) for the relative amplitude of the two characteristic offsets for polarized emission discussed in the text, $\left|\overline{\boldsymbol{\delta}\textbf{x}}_{\mathcal{P}}\right|$ and $\left|\overline{\boldsymbol{\delta} \textbf{x}}_{\left|\mathcal{P}\right|}\right|$, for images with $N=1$, 2, 5, and 10 patches of random placement and polarization. Each curve corresponds to the empirical CDF for $10^6$ realizations in which each of the $N$ patches had its location and polarization randomly drawn from the unit disk.
}
\label{fig::centroid_cdf}
\end{figure}

While this interpretation is exact and intuitive when the polarization has a uniform direction, it must be treated with care when the polarization direction varies. Specifically, $\mathcal{P}(\textbf{x})$ is complex so it cannot be thought of as a ``distribution'' of polarized flux with positive density. Indeed, in general, its associated ``centroid'' is complex, so the linear change in baseline can also affect the amplitude of the fractional polarization. Nevertheless, the most pathological cases occur when the image-averaged (i.e., the zero-baseline) polarization is especially small, and the complex centroid still tends to reflect more intuitive associations with the polarization image. 

\begin{figure*}[t]
\includegraphics*[width=1.0\textwidth]{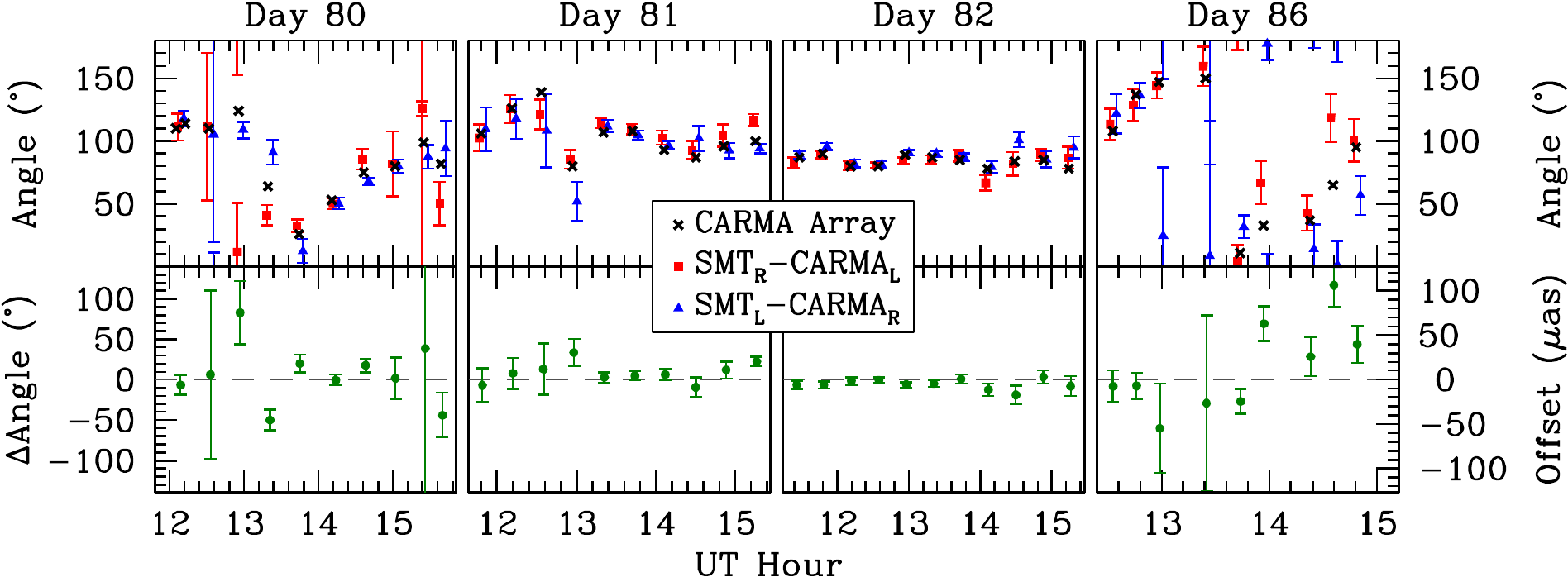} 
\caption{\small
{\bf Time-variable displacement of the polarization centroid of \sgra.} 
Polarization angle of \sgra\ seen on the two directions of the SMT-CARMA baseline and with CARMA (i.e., the EVPA) on four observing days; each triplet of simultaneous points (black, red, blue) is offset slightly in time for clarity. Plotted errors ($\pm1\sigma$) account for both thermal and calibration uncertainties. 
The difference of the two measurements on the SMT-CARMA baseline, $\Delta$Angle, provides a natural estimate of the offset between the centroids of the total and polarized flux (projected along the SMT-CARMA baseline direction) with a $1^{\circ}$ change in angle corresponding to a $0.9~\mu{\rm as}$ offset.
} 
\label{fig::SMT_CARMA_EVPA}
\end{figure*}

For instance, let $\overline{\boldsymbol{\delta}\textbf{x}}_{\mathcal{P}} \equiv \re \left[{\left \langle \textbf{x}\, \mathcal{I}(\textbf{x}) \right \rangle}/{\left \langle \mathcal{I}(\textbf{x}) \right \rangle} - \left \langle \textbf{x}\, \mathcal{P}(\textbf{x}) \right \rangle /\left \langle \mathcal{P}(\textbf{x})\right \rangle \right]$ denote the offset of the polarization centroid, as would be inferred by the phase rotation described by  Eq.~\ref{eq::m_leading}, and let $\overline{\boldsymbol{\delta} \textbf{x}}_{\left|\mathcal{P}\right|} \equiv {\left \langle \textbf{x}\, \mathcal{I}(\textbf{x}) \right \rangle}/{\left \langle \mathcal{I}(\textbf{x}) \right \rangle} - \left \langle \textbf{x}\, \left|\mathcal{P}(\textbf{x}) \right| \right \rangle /\left \langle \left| \mathcal{P}(\textbf{x})\right| \right \rangle$ denote the offset of the centroid of the amplitude of the polarization image at every point, which is perfectly well behaved. Then, consider $N$ independent pointlike patches of polarization,
 with locations and 
polarizations that are each drawn randomly from the unit disk. 
For $N=1$, we will always have $\overline{\boldsymbol{\delta}\textbf{x}}_{\mathcal{P}} = \overline{\boldsymbol{\delta} \textbf{x}}_{\left|\mathcal{P}\right|}$. For $N=2$, only $28\%$ of realizations have $\left|\overline{\boldsymbol{\delta}\textbf{x}}_{\mathcal{P}}\right| < \left|\overline{\boldsymbol{\delta} \textbf{x}}_{\left|\mathcal{P}\right|}\right|$, fewer than $1\%$ of realizations have $\left|\overline{\boldsymbol{\delta}\textbf{x}}_{\mathcal{P}}\right| < \frac{1}{2} \left|\overline{\boldsymbol{\delta} \textbf{x}}_{\left|\mathcal{P}\right|}\right|$, and $85\%$ of realizations give $\left|\overline{\boldsymbol{\delta}\textbf{x}}_{\mathcal{P}}\right|$ within a factor of two of $\left|\overline{\boldsymbol{\delta} \textbf{x}}_{\left|\mathcal{P}\right|}\right|$. Fig.~\ref{fig::centroid_cdf} shows the cumulative distribution function of $\left|\overline{\boldsymbol{\delta}\textbf{x}}_{\mathcal{P}}\right|/\left|\overline{\boldsymbol{\delta}\textbf{x}}_{\left|\mathcal{P}\right|}\right|$ for $N=$1, 2, 5, 
and 10. 
In general, variations in the polarization direction tend to increase the inferred offset of the polarization centroid, and higher order terms in Eq.~\ref{eq::m_leading} may destroy alignment in the phases of $\breve{m}(\pm \textbf{u})$ but are unlikely to introduce it. 
Thus, $\left| \overline{\boldsymbol{\delta}\textbf{x}}_{\mathcal{P}} \right|$ provides a reasonable estimate and reliable upper bound for the offset of the polarization centroid. 
Fig.~\ref{fig::SMT_CARMA_EVPA} shows the measured offsets of the polarization centroid for four days of our observations.

\paragraph*{Constraints on the Polarization Amplitude and the Degree of Order in the Fields.} Although $\breve{m}(\textbf{u})$ provides a robust observable, the examples of Fig.~\ref{fig::dictionary} show that it is difficult to precisely constrain the degree of order in the polarization field without also analyzing the behavior of the correlated total flux, $\tilde{\mathcal{I}}(\textbf{u})$. We now show that this pair of observables provides a very general quantification of the degree of order; in Fig.~\ref{fig::flux_fpol} of the main text, we compare our measurements of $\{ \tilde{\mathcal{I}}(\textbf{u}), \breve{m}(\textbf{u})\}$ with simple geometrical models and with a general relativistic magnetohydrodynamic (GRMHD) simulation. 

On long baselines, our highest fractional polarizations tend to coincide with the lowest normalized and deblurred visibilities for total flux (see Fig.~\ref{fig::CARMA_Long_Compare}). The simplest explanation for this trend is that our long baselines resolve more of the structure in the total flux than in the polarization -- a natural consequence of the variations in polarization direction that we have inferred from the asymmetry of $\breve{m}(\pm \textbf{u})$. 
To more precisely interpret this effect and understand its significance, we constructed a class of simple models with a constant, Gaussian distribution of the total flux but with a stochastic linear polarization field. For each model, the fractional polarization has constant amplitude across the image but the polarization direction varies, with a specified coherence length. We specified the polarization amplitude of each model such that the ensemble-average zero-baseline polarization was 5.2\%, to match the average of our CARMA measurements of \sgra. Because of this constraint, models with increased variations in the polarization direction (a shorter coherence length) also had an increased polarization fraction across the image. [Note that the zero-baseline polarizations of individual image realizations will vary randomly, so this model comparison would not be well motivated if the CARMA measurements did not show significant variation from day to day or even within individual epochs.]

To relate these models to our measurements, we used the models to construct synthetic data sets. To make a single data set, we first generated a separate random image for each day of our observation. Then, for each long-baseline measurement, we retained the baseline orientation but rescaled its length so that $\tilde{\mathcal{I}}(\textbf{u})$ for the model matched the observed value. We then calculated the fractional polarization on that baseline and replaced the observed value of the fractional polarization with the new value sampled for the model. Synthetic data sets generated in this way account for the covariance between measurements on the same observing track or at similar baseline lengths (e.g., SMA-CARMA and SMA-SMT). Because \sgra\ also shows time evolution within each observing day, this analysis likely overcompensates for the covariance, so true uncertainties may be smaller than what we infer. 

For 10{,}000 synthetic data sets generated with the coherence length used in case (\textbf{C}) in Fig.~\ref{fig::flux_fpol} (fields that are unresolved by current EHT baselines), only 4 yielded binned polarizations that were all below those that we measure, giving a $p$-value of $4\times10^{-4}$, equivalent to a significance of $3.4\sigma$. Likewise, for 100{,}000 synthetic data sets generated with the coherence length used in case (\textbf{A}) (highly ordered fields), none had samples with binned polarizations that were all higher than those we measured, giving a $p$-value ${\lsim}10^{-5}$, equivalent to a significance exceeding $4\sigma$. Our measurements thus favor a scenario in which the fields lie in an intermediate regime, with order that is 
partially 
resolved on our long baselines (an angular scale of ${\sim}60~\mu{\rm as}$). Using a Kolmogorov scaling to extrapolate the scattering kernel would lead to estimates with slightly more order in the fields. 

Numerical simulations \cite{Broderick_Loeb_2006,Dexter_2010,Shcherbakov_2012,Dexter_2014} indicate how physically motivated models may differ from these simple schematic models. A sample image from a GRMHD simulation is shown in Fig.~\ref{fig::flux_fpol} \cite{Gammie_2003,Shcherbakov_2012,Shcherbakov_2013}.  
In this image, the brightest regions have the lowest fractional polarizations. This result arises because the brightest regions are those with substantial Doppler-boosting -- they lie on the oncoming side of the accretion disk. This effect also increases the opacity of the emission region and thereby decreases the fractional polarization (we discuss this effect in detail below). In contrast, our schematic models have identical polarization fractions throughout the entire image, so their associated polarization fractions likely represent lower limits for the typical image polarization fractions. As Fig.~\ref{fig::flux_fpol} demonstrates, this GRMHD simulation exhibits a balance between order and variations in the fields that is compatible with our current measurements. Nevertheless, this GRMHD simulation is not fully compatible with our observations. While the simulation reproduces the spectrum of \sgra\ in the submillimeter bump and the linear polarization fraction seen at 230~GHz, the overall flux is too 
compact to reproduce our visibility amplitudes (Fig.~\ref{fig::amplitudes}). Additional work is needed to securely link our data to numerical simulations of \sgra.

\section*{Polarization Considerations for Sgr A*}

As noted in the main text, the 1.3-mm emission from \sgra\ is thought to be dominated by synchrotron emission from a relativistic thermal population of electrons near the black hole. To provide a point of contact with standard calculations in the literature, we now briefly consider expected polarization characteristics for synchrotron emission from a homogeneous slab of such material, and we derive the approximate dependence of fractional linear polarization $\Pi$ on optical depth $\tau$. [Note that this simplified one-zone model must be treated with caution for the emission near the black hole, which likely has steep temperature gradients.]

For a thermal population of emitting particles, the emission ($j$) and absorption ($\alpha$) coefficients in a basis $\{ \perp,\parallel \}$ determined by the projected magnetic field on the sky are related by Kirchoff's law: $j_{\parallel,\perp}/\alpha_{\parallel,\perp} = k T \left( \frac{\nu}{c} \right)^2$ \cite{RL}. Now, let $\tau \propto \frac{1}{2} \left( \alpha_{\parallel} + \alpha_{\perp} \right)$ denote the average optical depth, and let $\Pi_0 \equiv \Pi(\tau \rightarrow 0) = \frac{ j_{\perp} - j_{\parallel} }{ j_{\perp} + j_{\parallel}} = \frac{ \alpha_{\perp} - \alpha_{\parallel} }{ \alpha_{\perp} + \alpha_{\parallel}}$ define the linear polarization fraction in the optically-thin limit. For a uniform slab of material with linear depth $z$, the specific intensity takes the form $I_{\perp,\parallel} = \frac{j_{\perp,\parallel}}{\alpha_{\perp,\parallel}} \left( 1 - e^{-\alpha_{\perp,\parallel}z} \right)$ \cite{RL}. We can then write
\begin{align}
\Pi(\tau) = \frac{I_{\perp} - I_{\parallel}}{I_{\perp} + I_{\parallel}} = \left[ \frac{\mathrm{sinh}(\Pi_0 \tau)}{1 - \mathrm{cosh}(\Pi_0 \tau) e^{-\tau}} \right] e^{-\tau}.
\end{align}
Thus, the emission is unpolarized when the source is optically thick. 
This expression is a special case (a thermal emitting population) of Eq.~A3 in \cite{Homan_2009}, for instance. For large optical depths, $\tau \gg 1$, the polarization falls as $\Pi(\tau) \sim 0.5 e^{-\left(1-\Pi_0 \right) \tau}$.  

We can also estimate the polarization fraction in the optically-thin limit, $\Pi_0$:
\begin{align}
\Pi_0 \equiv \Pi(\tau \rightarrow 0) = \frac{ \int dE\, G(x) N(E) }{ \int dE\, F(x) N(E) }.
\end{align}
Here, $N(E)$ is the distribution function of particle energy, $F(x) \equiv x \int_{x}^{\infty} K_{\frac{5}{3}}(\xi) d\xi$, and $G(x) \equiv x K_{\frac{2}{3}}(x)$, where $x \equiv \nu/\nu_{\rm c}$ and $\nu_{\rm c} \equiv \frac{3 \gamma^2 e B \sin \varphi_{\rm p}}{4\pi m_{\rm e} c}$ defines a critical frequency for electrons of Lorentz factor $\gamma$ and pitch angle $\varphi_{\rm p}$ relative to a magnetic field of strength $B$ \cite{RL}. For relativistic thermal particles, $N(E) \propto E^2 e^{-E/kT}$, so we can rewrite the polarization as
\begin{align}
\Pi_0\left(\nu\right) =  \frac{ \int dx\, G(x) x^{-5/2} e^{-\sqrt{\hat{x}/x}} }{ \int dx\, F(x) x^{-5/2} e^{-\sqrt{\hat{x}/x}} }.
\end{align}
Here, $\hat{x} \equiv \nu/\hat{\nu}_{\rm c}$ and $\hat{\nu}_{\rm c} \equiv \nu_{\rm c}\left(\gamma = \theta_{\rm e}\right)$, where $\theta_{\rm e} = k T_{\rm e}/\left(m_{\rm e} c^2\right)$ is the dimensionless electron temperature. As $\hat{x} \rightarrow 0$ the polarization $\Pi_0 \rightarrow 1/2$, while as $\hat{x} \rightarrow \infty$ the polarization $\Pi_0 \rightarrow 1$. Note that these limits are identical for a monoenergetic distribution function \cite{Pacholczyk_1970}. 

We can further constrain $\Pi_0$ at 230~GHz using the measured spectrum of \sgra. Specifically, for optically-thin emission well above the critical frequency, the flux density falls slowly: $F_{\nu} \propto \frac{\nu}{\hat{\nu}_{\rm c}} e^{-1.8899 \left(\nu/\hat{\nu}_{\rm c}\right)^{1/3}}$ \cite{Mahadevan_1996}. However, the spectrum of \sgra\ falls steeply from the submillimeter to the near-infrared: $F_{230{\rm GHz}} \approx 3~{\rm Jy}$ and $F_{80{\rm THz}} \approx 5~{\rm mJy}$\ \cite{Schodel_2011}. This decrease in flux necessitates that $\hat{\nu}_{\rm c}\lsim 300~{\rm GHz}$ to avoid over-producing the quiescent near-infrared flux of \sgra. Hence, as a characteristic polarization for \sgra\ in the optically-thin limit at 230 GHz, we take $\hat{x}_{\rm 230GHz} \gsim 1$, giving $\Pi_0 \gsim 62\%$. 

We can then relate the measured polarization properties to the optical depth. For instance, the zero-baseline fractional polarization (up to ${\sim}$10\%) gives a lower-limit for the highest image polarization and requires a mean optical depth $\tau \lsim 4.5$. Our interferometric measurements sharpen this estimate through their estimate of the degree of disorder in the polarization field. For physically motivated models, our measurements suggest a characteristic image polarization of $\sim20-30\%$ (cf.~Fig.~\ref{fig::flux_fpol}), which requires a mean optical depth of $\tau \lsim 2.8-1.9$.

\begin{longtable}{cccccccccc}
\caption{{\bf Polarimetric detections of \sgra.} Calibrated polarimetric detections of \sgra\ in 2015. 
Each line gives the observing day, UT start time, scan duration, baseline code, band, baseline length, fractional polarization $\breve{m}$, and thermal noise $\sigma$ for the real and imaginary parts. 
Station codes are: D/E (CARMA Reference LCP/RCP), F/G (CARMA Phased LCP/RCP), J (JCMT RCP), P/Q (SMA LCP/RCP). To make the specified baseline $\{ u, v\}$ is consistent with the definition of $\breve{m}(u,v)$, the baseline direction, $\pm \{u, v\}$, depends on whether the cross-hand visibility is $\langle R_1 L_2^\ast \rangle$ or $\langle L_1 R_2^\ast \rangle$. [Data~S1 includes the complete dataset.]} \label{tab_pol} \\
\hline Day & Time & Length & Baseline & Band & $u$          & $v$           & Re$(\breve{m})$ & Im$(\breve{m})$ & $\sigma$ \\ 
           & (UT) & (s)    &          &      & (k$\lambda$) &  (k$\lambda$) &                 &                 &          \\ \hline\hline
\endfirsthead

\multicolumn{10}{c} 
{{\bfseries \tablename\ \thetable{}. Polarimetric detections of \sgra.}} \\
\hline Day & Time & Length & Baseline & Band & $u$          & $v$           & Re$(\breve{m})$ & Im$(\breve{m})$ & $\sigma$ \\ 
           & (UT) & (s)    &          &      & (k$\lambda$) &  (k$\lambda$) &                 &                 &          \\ \hline\hline
\endhead

\hline \hline
\endlastfoot

80 & 12:04 & 30 & SESD & low & 462734 & -79984 & -0.244 & -0.055 &  0.138\\
80 & 12:04 & 30 & SGSF & high & 462693 & -80014 &  0.043 & -0.062 &  0.058\\
80 & 12:04 & 30 & SGSF & low & 462693 & -80014 & -0.128 &  0.054 &  0.052\\
80 & 12:04 & 30 & TFTG & high & -462693 & 80014 &  0.007 & -0.047 &  0.064\\
80 & 12:04 & 30 & TFTG & low & -462693 & 80014 &  0.022 &  0.061 &  0.063\\
\multicolumn{9}{c}{\vdots}
\end{longtable}

\begin{longtable}{ccccccccc}
\caption{{\bf Normalized and deblurred visibilities of \sgra.} Detections of \sgra\ in 2015. Each line gives the observing day, scan start time, scan duration, baseline code (see Table~\ref{tab_pol}), band, vector baseline, normalized and deblurred visibility, and the aggregate of thermal and systematic noise. [Data~S2 includes the complete dataset.]} \label{tab_amp} \\
\hline Day & Time & Length & Baseline & Band & $u$          & $v$           & Visibility & $\sigma$ \\ 
           & (UT) & (s)    &          &      & (k$\lambda$) &  (k$\lambda$) &                 &          \\ \hline\hline
\endfirsthead

\multicolumn{9}{c}%
{{\bfseries \tablename\ \thetable{}. Normalized and deblurred visibilities of \sgra.}} \\
\hline Day & Time & Length & Baseline & Band & $u$          & $v$           & Visibility & $\sigma$ \\ 
           & (UT) & (s)    &          &      & (k$\lambda$) &  (k$\lambda$) &                 &          \\ \hline\hline
\endhead

\hline \hline
\endlastfoot

80 & 12:52 & 30 & QG & low & 2671921 & 1533107 &  0.199 &  0.031\\
80 & 12:53 & 480 & JG & high & 2675482 & 1527461 &  0.147 &  0.023\\
80 & 12:53 & 480 & JG & low & 2675482 & 1527461 &  0.134 &  0.021\\
80 & 12:53 & 480 & PF & low & 2675510 & 1527434 &  0.120 &  0.019\\
80 & 12:53 & 480 & SP & low & -3214510 & -1395150 &  0.140 &  0.026\\
\multicolumn{9}{c}{\vdots}
\end{longtable}

\end{document}